\documentclass[12pt]{article}
\usepackage{jheppub_modified_green}
\usepackage{graphicx}
\usepackage{epsfig}
\usepackage{amsfonts}
\usepackage{amssymb}
\usepackage{amsmath}
\usepackage{amsthm}
\usepackage{subfig}
\usepackage{bm}
\usepackage{hyperref}
\usepackage{mathrsfs}
\usepackage{bbm}
\usepackage{cancel}
\usepackage{color}
\usepackage{accents}

\usepackage[T1]{fontenc} 
\newtheorem{theorem}{\sf Theorem}

\newtheorem{conjecture}[theorem]{Conjecture}

\newcommand{\Res}{ \text{Res}}
\newcommand{\updown}[2]{{ \makebox [0 pt ]{$ {\scriptscriptstyle  { #1 \atop #2 } } $}  }}

\title{\boldmath A differential operator for integrating
one-loop scattering equations }

\author[a,b]{Gang Chen}
\author[b]{Yeuk-Kwan E. Cheung}
\author[b,1]{Tianheng Wang\note{Corresponding author.}}
\author[c]{Feng Xu}

\affiliation[a]{Department of Physics, Zhejiang Normal University, Jinhua, Zhejiang Province, China}
\affiliation[b]{Department of Physics, Nanjing University, Nanjing, Jiangsu Province, China}
\affiliation[c]{Weavi Corporation Limited, Nanjing, Jiangsu Province, China}

\emailAdd{gang.chern@gmail.com}
\emailAdd{cheung@nju.edu.cn}
\emailAdd{tianhengwang@outlook.com}
\emailAdd{schyfeng@gmail.com}

\abstract{We propose a differential operator for computing the residues associated with a class of meromorphic $n$-forms that frequently appear in the Cachazo-He-Yuan form of the scattering amplitudes. This differential operator  is conjectured to be uniquely determined by the local duality theorem and the intersection number of the divisors in the $n$-form. 
We use  the operator to evaluate  the tree-level amplitude 
of $\phi^3$ theory and the one-loop integrand of 
Yang-Mills theory from their CHY forms. 
The method can reduce the complexity of the calculation. 
In addition, the expression for the 1-loop four-point 
Yang-Mills integrand obtained in our approach has a clear correspondence with the Q-cut results.
}

\begin{document} 
\maketitle
\flushbottom

\section{Introduction}
\label{sec:intro}

Cachazo, He and Yuan discovered a succinct form of  
writing the scattering amplitudes of  various (quantum) field 
theories~\cite{Cachazo:2013hca, Cachazo:2013iea, 
Cachazo:2014nsa}. 
Not only  does it make many duality properties manifest, e.g.  
the KLT relation~\cite{Kawai:1985xq} between gauge theory and 
gravity~\cite{Cachazo:2013gna} as well as the KK relation~\cite{Kleiss:1988ne}
 and the  BCJ relation~\cite{Bern:2008qj}, 
  but it is also a great platform for constructing 
new theories, with or without Lagrangian 
descriptions~\cite{Cachazo:2014xea}.  
A flurry of activities  ensued.

In four dimensions, tree-level scattering equations can be viewed as the constraints of the Roiban-Spradlin-Volovich (RSV) formula in $\mathcal N=4$ super Yang-Mills~\cite{Roiban:2004yf, Witten:2003nn}. The $\mathcal N=8$ supergravity tree amplitudes, 
proposed and derived from twistor string theory~\cite{Cachazo:2012pz, Skinner:2013xp}, can also be included in this 
framework~\cite {Cachazo:2013iaa,Cachazo:2012da,Cachazo:2012kg}. 
In $d$ dimensions, the  polynomial form of the scattering equations is first obtained  in~\cite{Dolan:2014ega}. 
The algebraic varieties corresponding to these homogeneous polynomials 
are studied by~\cite{He:2014wua}.

A few solutions of the tree-level  scattering 
equations in four dimensions  in some special kinematic 
limits  are obtained:
tree level scattering amplitudes for $n$ gluons and $n$ gravitons are computed  in~\cite{Kalousios:2013eca},
the tree-level scattering equations are solved  up to 
six-points~\cite{Weinzierl:2014vwa}, while a
 general  relation between solutions of the tree-level scattering equations is proposed,  and checked at special kinematic limits up to six points~\cite{Lam:2014tga}. 
 
MHV tree-level amplitudes of gravity and gluons are obtained from the CHY amplitudes in~\cite{Du:2016blz}. 
A five-point scattering amplitude in  Yang-Mills theory  is obtained in~\cite{Kalousios:2015fya} by exploiting the 
Vieta formula (relating the sum of the solutions of a polynomial equation to the coefficients of the polynomial).  
The elimination theory  is  applied  to 
reduce  the polynomials and obtained 
the residues of the scattering equations 
 in~\cite{Cardona:2015eba, Cardona:2015ouc, Dolan:2015iln}.

 Methods based on algebraic geometry are exploited to improve 
 the efficiency of computations. 
 Companion matrix method~\cite{Huang:2015yka} 
and Bezoutian matrix method~\cite{Sogaard:2015dba} are used to evaluate the CHY expressions, without solving the scattering equations, and checked against the 5-point amplitude in $\phi^3$ theory analytically, as well as higher point amplitudes numerically. 
In~\cite{Bosma:2016ttj}  polynomial reduction techniques are 
used to cast  the scattering equations into the standard basis,  called the H-basis. 
In~\cite{Zlotnikov:2016wtk} another prescription for such evaluation is proposed, using the polynomial reduction techniques, and explicit results are given analytically up to 3-points at one loop in $\phi^3$ theory. 
These approaches do not depend on the particular theory and the method we are going to propose also belongs to this category.

Building blocks method  was proposed  
by~\cite{Cachazo:2015nwa} in which higher-point CHY-integrands
 are reduced a la BCFW to basic building blocks.
It was then further developed in~\cite{Gomez:2016bmv} to a systematic $\Lambda$-algorithm, and then used in~\cite{Cardona:2016bpi} to propose and compute a 1-loop  CHY amplitude for the $n$-gon. Integration rules for higher-point amplitudes are  derived 
in~\cite{Baadsgaard:2015voa, 
Lam:2015sqb,  Lam:2016tlk, Baadsgaard:2015hia, 
Mafra:2016ltu, Huang:2016zzb} to facilitate practical computations. High order poles were discussed 
in~\cite{Bjerrum-Bohr:2016juj, Cardona:2016gon}. 

In~\cite{Cachazo:2015aol} one loop scattering amplitudes 
are obtained from  tree-level ones in one higher 
spatial dimension.
A universal all-loop CHY form was constructed from tree-level CHY forms in $\phi^3$ theory in~\cite{Feng:2016nrf} and checked against Q-cut results. 

These methods have various degrees of 
success in constructing higher-loop amplitudes 
 in scalar theories as successful methods have been developed 
 to subtract the forward singularities arisen in gluing tree-level amplitudes  to form one-loop amplitudes. 
These  methods can hence be generalized in a straightforward way to the computations of scattering  amplitudes in gauge theory 
and gravity at tree-level;
but so far no general method is in sight for removing the forward singularities introduced when gluing tree amplitudes  to form loop  amplitudes in gauge or gravity theories.  

On the other hand the  scattering equations at loop levels are  derived from ambitwistor string theory~\cite{Berkovits:2013xba, Mason:2013sva, Geyer:2014fka}. 
The CHY expressions are then  extended  to one 
and two loops for the bi-adjoint scalar, gauge theory, and gravity in~\cite{Geyer:2015bja, Geyer:2015jch, Geyer:2016wjx}. 

In this paper, we propose a new method, also based on algebraic 
geometry, to evaluate CHY forms. 
A differential operator, constructed in a systematic procedure, is conjectured to capture the residue around the contour associated with the scattering equations.
A one-to-one correspondence to the Q-cut results for the 
1-loop four-point CHY expression in SYM is made manifest 
by our algorithm.
Compared with other algebraic geometry based techniques, 
we avoids the task of finding  the Gr{\"o}bner basis~\cite{cox2006using} of the scattering equations. 
The construction of this operator demands information mostly 
from the integration contour; and therefore the complexity
 of the theory-specific factors in CHY forms has little 
 impact on the procedure.
The residues at the phantom poles resulted from
factorization in  the 
polynomial form of scattering equations  naturally 
vanish in our method.

\section{A differential operator for multivariable residues}
\label{sec:operator}
The CHY-form provides a beautiful and compact expression for  scattering amplitudes. 
Schematically, an $n$-particle scattering process reads
\begin{eqnarray}
 \int {d\sigma_1\cdots d\sigma_{n}\over \text{vol}(\text{Residual Symmetry}) } \delta(\text{Scattering Equations}) \,\mathcal{I},  
 \label{CHY}
\end{eqnarray}
with  $\sigma_i$'s being complex variables which can, in turn, 
be related to the worldsheet coordinates of the vertex operators in string theory. The integrand $\mathcal I$ depends on the underlying theory. These scattering equations are typically a 
set of rational equations in  $\sigma_i$'s, whose coefficients encoding the dynamic information of the scattering process. 

In the language of complex analysis,
 integrals with delta-functions are equivalent to  residues 
 around the contours  defined by these delta-function constraints. 
The CHY-form~(\ref{CHY}) can be casted into a residue associated with a meromorphic form as follows,
\begin{eqnarray}\label{eq:chy2}
\oint \frac{d\sigma_1\wedge \cdots \wedge d\sigma_{n-m}}{ \, h_1\cdots h_{n-m}} \,\mathcal I'\,,
\end{eqnarray}
where $m$ is the number of the residual symmetry generators and $h_i$'s are polynomials originated from the scattering 
equations. These polynomials are, roughly speaking, the 
numerators in the polynomial scattering equations. 

In this section, we introduce a differential operator for computing such residues.
\begin{conjecture}\label{conj:1}: Given a polynomial ideal $\langle f_1,f_2\cdots f_{k}\rangle$ in variables $z_1,z_2\cdots z_{k}$,  the polynomials are  homogeneous and their degrees are  $d_1,d_2,\cdots, d_{k}$ respectively.  If the solution to the corresponding algebraic equations is an isolated point $p$, the residue associated with a meromorphic form has a differential interpretation. Namely, for a holomorphic function $\mathcal R(z_i)$ in the neighbourhood of the point $p$, 
\begin{eqnarray}
\Res_{\{(f_1),\cdots,(f_k)\},p}[\mathcal{R}]\equiv\oint \frac{dz_1\wedge \cdots \wedge dz_k}{f_1\cdots f_k} \mathcal R =\mathbb{D}[\mathcal R], ~~~\mathbb{D}=\sum_{\{s_i\}}a_{s_1 s_2 \cdots s_{k}}\partial^{s_{1}}_{r_1}\partial^{s_2}_{r_2}\cdots \partial^{s_{k}}_{r_{k}}\,,~~~ ~\label{defD}
\end{eqnarray}
where the coefficients $a_{s_1s_2 \cdots s_{k}}$ are constants independent of $z_i$ and $
\partial^{s_i}_{r_i}={\partial^{s_i}\over \partial z^{s_i}_{r_i}}\,,  i\in[1,k]$.
The summation is taken over all the solutions to the Frobenius equation:
\begin{eqnarray}
\sum_{i=1}^{k} s_i= \sum_{h=1}^k d_h -k~. 
\end{eqnarray}
Furthermore, the differential operator $\mathbb D$ is uniquely determined by requiring the residue to satisfy the local duality theorem~\cite{hartshorne2013algebraic, griffiths2014principles} and to give the correct intersection number of the 
divisors $D_i=(f_i)$.~\footnote{We remark on the motivation 
for  this conjecture. Let  us consider the integral,
\begin{eqnarray}
\oint\limits_{f_1=\cdots f_k=0} \frac{g(z_1,\cdots, z_k)\, dz_1\wedge \cdots \wedge dz_k}{f_1 \cdots f_k}\,,\nonumber
\end{eqnarray}
where $f_i$'s are homogeneous polynomials in $z_i$'s and 
the numerator $g$ is a monomial in $z_i$ of degree $M$. 
The residue is non-vanishing if and only if
 $M=\sum_{i=1}^k \, d_{f_i}$~\cite{Sogaard:2015dba}. 
 Therefore, the differential operator that computes the residue  can involve only derivatives of degree $M$. 
 This observation generalizes naturally to the case in which  
 the numerator is a polynomial. 
}
\end{conjecture}
We shall exploit this conjecture to evaluate CHY scattering equation~(\ref{eq:chy2}).
 Typically, the polynomials $h_i$ in~(\ref{eq:chy2}) are not homogeneous, and therefore  include extra poles that are not solutions to the original scattering equations. 
 We call these ``spurious poles,''  and the locations of them 
 are easy to determine. 
 In Section~\ref{sec:ansatz}, we shall introduce a ``homogenization''  procedure to case  these polynomials and the resulting integrals to  meet the conditions of the conjecture.
 
The  differential operator $\mathbb D$ computes the sum of residues around all the solutions of $h_1 = \cdots = h_{n-m}=0$, including the spurious ones.  
Therefore it is crucial to  demonstrate how to remove the 
contributions from the spurious poles. 
A calculation of CHY amplitudes can  sometimes  involve
 computing the residues at infinity. 
This is achieved using our conjecture, as demonstrated 
 in Section~\ref{sec:ansatz}.
By computing residues at the finite poles, the spurious poles 
and the poles at infinity,   scattering amplitude can be conveniently evaluated from the CHY forms. 


\section{A Warm-up example}
\label{sec:example}
In this section  we study the 5-point tree-level amplitude in the massless $\phi^3$ theory and use  it as a toy model to illustrate
 the evaluation of the multi-variable contour integrals that often appear in CHY-form by the proposed operator $\mathbb D$.
We denote this amplitude as $\mathcal A_{\phi^3}$ and its explicit expression is given in~\cite{Dolan:2014ega,Huang:2015yka,Sogaard:2015dba},
\begin{equation}
\label{eq:FiveTree}
 \mathcal A_{\phi^3}=\oint\limits_{h_1=h_{2}=0} \frac{d\sigma_1 \wedge d\sigma_2}{{h}_1 {h}_2 (\sigma_1-\sigma_2)}+\frac{d\sigma_1 \wedge d\sigma_2}{{h}_1 {h}_2 (1-\sigma_1) \sigma_2}\,,
\end{equation}
where ${h}_1, {h}_2 $  being   polynomials in $\sigma_1$ and $\sigma_2$, are roots of the tree-level scattering equations. 
Their solutions  take the following forms, 
\begin{eqnarray}
h_1&=&k_{13} \sigma_1+k_{14} \sigma_2+k_{12} \, ,\\
h_2&=&k_{45} \sigma_1 +k_{25} \sigma_1 \sigma_2 +k_{35} \sigma_2 \,,
\end{eqnarray}
where $k_{i
j}={1\over 2}(k_{i}+ 
k_{j})^2$. 

The two terms in~(\ref{eq:FiveTree}) can be integrated separately.
In the first term  we denote the remaining factor in the denominator as $h_0 = (\sigma_1-\sigma_2)$. 
These three polynomials $h_1, h_2, h_0$ can be made homogeneous:
\begin{eqnarray*}
\tilde h_0&=&(\sigma_1-\sigma_2) \,, \\
\tilde h_1&=&k_{13} \sigma_1+k_{14} \sigma_2+k_{12}\sigma_0 \,, \\
\tilde h_2&=&k_{45} \sigma_0\sigma_1 +k_{25} \sigma_1 \sigma_2 +k_{35} \sigma_0\sigma_2 \,,
\end{eqnarray*}
by introducing  an extra variable $\sigma_0$
which will be later integrated out.  
Hence the first contour integral becomes,
\begin{eqnarray}
\oint\limits_{h_1=h_{2}=0} \frac{d\sigma_1 \wedge d\sigma_2}{{h}_1 {h}_2 (\sigma_1-\sigma_2)} &=& \oint\limits_{{\tilde h}_1= {\tilde h}_{2}= \sigma_0-1=0} {d\sigma_1\wedge d\sigma_{2}\wedge d\sigma_0 \over  {\tilde h}_1{\tilde h}_{2} (\sigma_0-1)} {1\over \tilde h_0} \nonumber\\
& =&-\oint\limits_{{\tilde h}_1= {\tilde h}_{2} = \tilde h_0=0} {d\sigma_1\wedge d\sigma_{2}\wedge d\sigma_0 \over  {\tilde h}_1 {\tilde h}_{2} {\tilde h}_0} {1\over (\sigma_0-1)}\, ,
\end{eqnarray}
where in the second equality  sign we have used the global residue theorem and the intersecting divisors are all generated by homogeneous polynomials. This contour integral is immediately computed by our conjecture. The corresponding differential operator takes the form, 
\begin{equation}
\mathbb{D}=a_{100}{\partial\over\partial \sigma_1}+a_{010}{\partial\over\partial \sigma_2}+a_{001}{\partial\over\partial \sigma_0}. 
\end{equation}
The computation of the integral is now translated to finding the values of the coefficients $a_{100}$, $a_{010}$ and $a_{001}$. The local duality theorem requires, 
 \begin{eqnarray}
\oint\limits_{{\tilde h}_1= {\tilde h}_{2} = \tilde h_0=0} { {\tilde h}_1d\sigma_1\wedge d\sigma_{2}\wedge d\sigma_0 \over  {\tilde h}_1 {\tilde h}_{2} {\tilde h}_0} =0\,, \quad\quad
\oint\limits_{{\tilde h}_1= {\tilde h}_{2} = \tilde h_0=0} { {\tilde h}_0 d\sigma_1\wedge d\sigma_{2}\wedge d\sigma_0 \over  {\tilde h}_1 {\tilde h}_{2} {\tilde h}_0} =0 \, .
\label{localdual}
\end{eqnarray}
The intersection number of the divisors here is 2 and this yields, 
 \begin{eqnarray}
\oint\limits_{{\tilde h}_1= {\tilde h}_{2} =\tilde h_0=0} {  d {\tilde h}_1\wedge d {\tilde h}_2\wedge d {\tilde h}_0 \over  {\tilde h}_1 {\tilde h}_{2} {\tilde h}_0} =\oint\limits_{{\tilde h}_1= {\tilde h}_{2} = \tilde h_0=0} { \mathcal{J} d {\sigma}_1\wedge d {\sigma}_2\wedge d {\sigma}_0 \over  {\tilde h}_1 {\tilde h}_{2} {\tilde h}_0} =2\,, \label{internum}
\end{eqnarray}
where $\mathcal{J}=\det({\partial \tilde h_i\over \partial \sigma_j})$ is the Jacobian of integral parameter transformation. In the language of the differential operator, conditions (\ref{localdual}) and (\ref{internum}) read,
\begin{eqnarray}
\label{eq:FiveLocalDuality}
\mathbb{D} {\tilde h}_1 =0 \,, \quad\quad
\mathbb{D} {\tilde h}_0=0 \,, \quad\quad 
\mathbb{D} \mathcal{J} =2 \,.
\end{eqnarray}
Solving for $a$'s the constraints above, we obtain,
\begin{eqnarray*}
a_{100}=-{2k_{12}\over G_1} \,, \quad\quad
a_{010}=-{2k_{12}\over G_1} \,, \quad\quad
a_{001}=-{2(k_{13}+k_{14})\over G_1} \,.
\end{eqnarray*}
where
\begin{eqnarray*}
G_1=-2 k_{12}^2 k_{134}+2 k_{13} k_{12} k_{123}+2 k_{14} k_{12} k_{123}+2 k_{13} k_{12} k_{124}+2 k_{14} k_{12} k_{124}\,.
\end{eqnarray*}
Thus the action of the differential operator gives,
\begin{equation}
\oint\limits_{{\tilde h}_0= {\tilde h}_{2} = \tilde h_0=0} {d\sigma_1\wedge d\sigma_{2}\wedge d\sigma_0 \over  {\tilde h}_1 {\tilde h}_{2} {\tilde h}_0} {1\over (\sigma_0-1)} =
\mathbb{D}\left(\frac{1}{\sigma_0-1}\right)=-\frac{a_{001}}{(\sigma_0-1)^2}\lvert_{\sigma_0\rightarrow 0}={2(k_{13}+k_{14})\over G_1}.
\end{equation}
Similarly, the second term in (\ref{eq:FiveTree}) can also be related to another integral in which the intersecting divisors are originated from homogeneous polynomials only. The residue of the latter is then represented by a second order differential operator $\mathbb D$,   
\begin{eqnarray}
\mathbb{D}=a_{002}{\partial^2\over \partial \sigma^2_0}+a_{011}{\partial\over \partial \sigma_0}{\partial\over \partial \sigma_2}+a_{020}{\partial^2\over \partial \sigma^2_2}+a_{101}{\partial\over \partial \sigma_1}{\partial\over \partial \sigma_0}+a_{110}{\partial\over \partial \sigma_1}{\partial\over \partial \sigma_2}+a_{200}{\partial^2\over \partial \sigma^2_1}\,.\nonumber
\end{eqnarray}
The residue computed by this operator is,
\begin{eqnarray*}
&& \mathbb{D} \left(\frac{1}{\sigma_0-1}\right)=-{32 k_{13} k_{14} (k_{13} (k_{124}+k_{134})-k_{14} k_{123})\over G_2}\,, \\
&& G_2=32 k_{12} k_{13} k_{14} k_{123} (-k_{14} k_{123}+k_{12} k_{124}+k_{13} k_{124}+k_{12} k_{134}+k_{13} k_{134})\,.
\end{eqnarray*}
Putting together the two terms, we obtain,
\begin{eqnarray}
\mathcal A_{\phi^3} = -{2(k_{13}+k_{14})\over G_1}+{32 k_{13} k_{14} (k_{13} (k_{124}+k_{134})-k_{14} k_{123})\over G_2}\,.
\end{eqnarray}
The expression is identical with those in  \cite{Dolan:2014ega,Huang:2015yka,Sogaard:2015dba}.

\section{Direct evaluation of one-loop CHY-form}
\label{sec:ansatz}
In this section we provide a systematic algorithm that exploits the conjecture (\ref{conj:1}) in the calculation of CHY forms. 

Before discussing the technical details, let us first sketch out the steps in words. The polynomial form of scattering equations is our starting point. The general transformations from the standard scattering equations to the polynomial ones are given in~\cite{Dolan:2014ega}. These transformations introduce a Jacobian into the CHY expression that is simply the Vandermonde determinant. The polynomial equations are not entirely equivalent to the original equations, rather, they bring in extra solutions that do not satisfy the original. 

As explained in Section \ref{sec:operator}, the evaluation of a CHY-form boils down to computing the residue (\ref{eq:chy2}). Since the polynomial equations have extra solutions, (\ref{eq:chy2}) can be obtained by computing the sum of the residues at \textit{all} the poles of said polynomials first and then removing the contribution from the extra poles.
To calculate these residues directly can be quite demanding and this is where Conjecture \ref{conj:1} comes in handy. 

Depending on the specific expression of the CHY-form, we may encounter two kinds of meromorphic forms: one that is regular at infinity and one that has non-vanishing residues at infinity. To compute the total residue in the former category, we adopt a straightforward procedure of homogenizing the polynomials such that all the poles are condensed at one single isolated point. Hence the total residue can be immediately determined by our conjecture. The latter category can be attacked in a similar fashion, after we embed the complex manifold on which the corresponding meromorphic form lives into a compact one and make a point at infinity well-defined. This process will be discussed in detail. 

As for the residues at the phantom poles, we observe that these poles are trivial to locate, however, for reasons that will become clear later, the aforementioned homogenization does not work for this case. A modified conjecture will be given for computing such residues.

\subsection{Polynomial scattering equations}\label{sec:polynomialSE}
An $n$-point scattering amplitude in $D$ dimensions takes the form,
\begin{eqnarray}
\mathcal A_n^{l=1} = \int \frac{d^D \ell}{\ell^2}\, \mathcal I^{l=1}_n ( \text{kinematics} )\,,
\end{eqnarray}
where $\ell$ denotes the loop momentum and the exact expression for the integrand $\mathcal I^{l=1}_n$ depends on the underlying quantum field theory. As shown in \cite{Geyer:2015bja, Geyer:2015jch, Geyer:2016wjx}, in a variety of theories, the integrand has a CHY representation that schematically reads,
\begin{equation}\label{eq:oneloopInt}
 \mathcal I_n^{l=1}=\oint\limits_{f_1=\cdots=f_{n-1}=0} {d\sigma_1\wedge\cdots\wedge d\sigma_{n-1} \over  f_1\cdots f_{n-1}} \frac{\mathcal N(\sigma_i)}{\mathcal D(\sigma_i)}\, ,
\end{equation}
where $\mathcal N(\sigma_i)$ and $\mathcal D(\sigma_i)$ are polynomials in $\sigma_i$'s and encode the kinematic information of the scattering amplitude.
The rational function $f_i$ denotes the $i$-th one-loop scattering equation, originally derived in the context of the high-energy limit of string theory in \cite{Gross:1989ge}, and later re-discovered in ambitwistor string in \cite{Berkovits:2013xba, Mason:2013sva},
\begin{equation}
f_i={\ell\cdot k_i \over \sigma_i} +\sum_{j\neq i}{k_i\cdot k_j \over \sigma_i-\sigma_j} = 0\, .
\end{equation}

While these equations are difficult to solve analytically, Dolan \textit{et al} introduced in \cite{Dolan:2014ega} the following transformations to make them more friendly,
\begin{equation}
 g_m=\sum_{i=1}^{n}\sigma^{m+1}_if_i\, , \quad\quad\quad m\in\{-1,0,\cdots\} \,. 
\end{equation}
By counting the degrees of freedom, obviouly only $n-1$ equations are independent. We choose the $(n-1)$ ones with the lowest degrees, that is, $g_i=0\; (i=1,\cdots n-1)$.\footnote{ $g_{-1}$ is still a rational function and $g_0=0$ is satisfied trivially by the conservation of momentum.} The transformations from $\{f_1,\cdots f_{n-1}\}$ to $\{g_1,\cdots, g_{n-1}\}$ bring into the integrand the following Jacobian that can be easily computed \cite{Dolan:2014ega},
\begin{eqnarray}
\mathcal J &=& \det \left(
\begin{array}{cccc}
\sigma_1 (\sigma_1-\sigma_n)\quad & \sigma_2 (\sigma_2 -\sigma_n)\quad & \cdots & \sigma_{n-1} (\sigma_{n-1}-\sigma_n)\\
\sigma_1 (\sigma_1^2-\sigma_n^2)\quad & \sigma_2 (\sigma_2^2-\sigma_n^2)\quad& \cdots & \sigma_{n-1} (\sigma_{n-1}^2-\sigma_n^2) \\
\vdots & \vdots & \ddots & \vdots \\
\sigma_1 (\sigma_1^{n-1}-\sigma_n^{n-1})\quad & \sigma_2 (\sigma_2^{n-1}-\sigma_n^{n-1})\quad & \cdots &\sigma_{n-1} (\sigma_{n-1}^{n-1}-\sigma^{n-1}_n)
\end{array}
\right)\nonumber\\
&=& \sigma_1\sigma_2 \cdots \sigma_{n-1} \prod_{1<i<j<n} (\sigma_i-\sigma_j)\,.
\end{eqnarray}
The explicit expressions for these new equations read,
\begin{eqnarray}
g_m = \sum_{i=1}^{n} p\cdot k_i \sigma_i^m + \sum_{i<j}^n k_i\cdot k_j \left(\sum_{r=1}^{m-1} \sigma_i^r \sigma_j^{m-r}\right)=0 \,,\quad\quad m=1,2,\cdots, n-1\,.
\end{eqnarray}
These polynomial equations already provide us a much better platform than $f_i$ since each of them is of degree $m$ while all the $f_i$'s give rise to equations of degree $n-2$ (with the choice of gauge $\sigma_n=1$). We can further simplify the equations by applying a few more linear transformations given as follows,
\begin{eqnarray}\label{eqn:hdef}
h_1&=&g_1\, ,\nonumber\\
h_2&=&g_2-g_1 (\sum_i^n \sigma_i) \,, \nonumber\\
h_3&=&g_3-g_2 (\sum_i^n \sigma_i)+{g_1\over 2}(\sum_{i\neq j}^n\sigma_i\sigma_j) \,, \nonumber\\
&\cdots&\nonumber\\
h_{n-1}&=&g_{n-1}-g_{n-2} (\sum_i^n \sigma_i)+{g_{n-3}\over 2}(\sum_{i_1\neq i_2}^n \sigma_{i_1}\sigma_{i_2})+\cdots\nonumber\\
&+&{g_{n-1-m}\over (-)^m m!}(\sum_{i_1\neq\cdots\neq i_m}^n\prod_{r=1}^m\sigma_{i_r})
+\cdots+{(-)^{n-2} g_{1}\over (n-2)!}(\sum_{i_1\neq\cdots\neq i_{n-2}}^n\prod_{r=1}^{n-2}\sigma_{i_r}).
\end{eqnarray}
The Jacobian for the additional transformations above is simply $1$. Explicitly, we write down the $h_i$'s,
\begin{eqnarray}
h_1&=&\sum_{i=1}^n l_i \sigma_i\nonumber\\
h_m&=&(-)^{m-1}\sum_{i_1<i_2\cdots<i_m}^n \sigma_{i_1\cdots i_m} l_{i_1\cdots i_m}
\label{eq:polynomialScatEq}
\end{eqnarray}
where $m\in \{2,\cdots,n-1\}$ and we have used the following notations,
\begin{equation*}
\sigma_{i_1\cdots i_m}\equiv \prod_{r=1}^m\sigma_{r}\,,\quad l_i\equiv l\cdot k_i\,, \quad l_{i_1\cdots i_m}\equiv \left(l\cdot k_{i_1\cdots i_m}{ -\frac{1}{2}} k_{i_1\cdots i_m}^2\right), \quad k_{i_1\cdots i_m}\equiv\sum_{r=1}^m k_{i_r}\,. 
\end{equation*}

As mentioned before, the polynomial equations $h_i=0\, (i=1,\cdots n-1)$ have more solutions than the original scattering equations. These extra solutions locate at  $\sigma_1 = \sigma_2 =\cdots = \sigma_{n-1}=1$ in the gauge $\sigma_n=1$.

Now the one-loop CHY-form (\ref{eq:oneloopInt}) can be rewritten as the following
\begin{eqnarray}\label{rewritten1loopAwithh}
 && \mathcal I_n^{l=1} \nonumber\\
& = &\oint\limits_{h_1=\cdots=h_{n-1}=0} {d\sigma_1\wedge\cdots\wedge d\sigma_{n-1} \over  h_1\cdots h_{n-1}}  {\mathcal{N'}(\sigma_i) \over \mathcal D(\sigma_i)} -\oint\limits_{\sigma_1=\cdots \sigma_{n-1}=1} {d\sigma_1\wedge\cdots\wedge d\sigma_{n-1} \over  h_1\cdots h_{n-1}} {\mathcal{N'}(\sigma_i) \over \mathcal D(\sigma_i)} \nonumber\\
&=& \oint\limits_{h_1=\cdots=h_{n-1}=0}  {d\sigma_1\wedge\cdots\wedge d\sigma_{n-1} \over  h_1\cdots h_{n-1}}  {\mathcal{N}_{reg}(\sigma_i) \over \mathcal D(\sigma_i)} +\oint\limits_{h_1=\cdots=h_{n-1}=0} {d\sigma_1\wedge\cdots\wedge d\sigma_{n-1} \over  h_1\cdots h_{n-1}}  {\mathcal{N}_{\infty}(\sigma_i) \over \mathcal D(\sigma_i)}  \nonumber\\
&& \quad -\oint\limits_{\sigma_1=\cdots \sigma_{n-1}=1} {d\sigma_1\wedge\cdots\wedge d\sigma_{n-1} \over  h_1\cdots h_{n-1}} {\mathcal{N'}(\sigma_i) \over \mathcal D(\sigma_i)} \,,
\end{eqnarray}
where $\mathcal N'(\sigma_i) = \mathcal J(\sigma_i) \mathcal N(\sigma_i) = \mathcal N_{reg}+\mathcal N_{\infty}$. In the second equal sign we have separated the numerator $\mathcal N'$ into two parts, with
\begin{align}
& \text{deg}(\mathcal N_{reg}) < \text{deg} (h_1) +\cdots \text{deg} (h_{n-1}) +\text{deg}(\mathcal D) - (n-1)\,,\\
& \text{deg}(\mathcal N_{\infty}) \geq \text{deg} (h_1) +\cdots \text{deg} (h_{n-1}) +\text{deg}(\mathcal D) - (n-1)\,.
\end{align}
The integrand containing $\mathcal N_{reg}$ has no residues at infinity while the one containing $\mathcal N_{\infty}$ does. In the rest of this section, we calculate these three terms in (\ref{rewritten1loopAwithh}), using our conjecture.

\subsection{Residues at finite poles}
\label{eq:finitepole}
In this subsection, we demonstrate how to utilize Conjecture \ref{conj:1} in computing the first term in the second equal sign of (\ref{rewritten1loopAwithh}). This term is equal to the sum of residues associated with the $n$-form regular at infinity:
\begin{eqnarray}
\Omega = {d\sigma_1\wedge\cdots\wedge d\sigma_{n-1} \over  h_1\cdots h_{n-1}}  {\mathcal{N}_{reg}(\sigma_i) \over \mathcal D(\sigma_i)} \,.
\end{eqnarray}
This residue can not be evaluated by the conjecture yet, but will be after the so-called ``homogenization'' procedure.


Consider one of the polynomials $h_i$ that is of degree $d$ and reads,
\begin{eqnarray}
h_i = \sum_{\{s\}} \alpha_{s_1 \cdots s_{n-1}} \sigma^{s_1}_1 \cdots \sigma^{s_{n-1}}_{n-1}\,, 
\end{eqnarray}
where $\alpha$'s are constants of which the explicit expressions are not important and we have $0\leq s_i \leq d$ for $i=1,\cdots n-1$ and $s_1+\cdots s_{n-1}\leq d$ for every term in the summation.
We introduce an additional variable $\sigma_0$ and define the homogenization of $h_i$ as the following,
\begin{eqnarray}
\tilde h_i = \sum_{\{s\}} \alpha_{s_1 \cdots s_{n-1}} \sigma^{s_1}_1 \cdots \sigma^{s_{n-1}}_{n-1} \sigma_0^{d-s_1-\cdots -s_{n-1}}\,.
\label{eq:Homgen}
\end{eqnarray}
When $\sigma_0 =1$ we recover the original polynomial $\tilde h_i=h_i $. We homogenize all the polynomials $h_i, ~(i=1,\cdots n-1)$ as well as $h_0 = \mathcal D(\sigma_i)$. We construct a new meromorphic form from these homogenized polynomials as follows,
\begin{eqnarray}
\tilde\Omega = \frac{\mathcal N_{reg}(\sigma_i)}{\tilde h_1 \cdots \tilde h_{n-1}} \frac{1}{\tilde h_0~(\sigma_0-1)} d\sigma_1\wedge\cdots\wedge d\sigma_{n-1}\wedge d\sigma_0\,.
\end{eqnarray}
Since the function in $\tilde\Omega$ is regular at infinity, the global residue theorem leads to,
\begin{eqnarray}
0 = \sum_p \Res_{ \{ D_1,\cdots, D_{n-1}, \sigma_0-1\}, p} ~\tilde\Omega+ \Res_{ \{D_1,\cdots, D_{n-1}, D_0\}, p'}~\tilde\Omega \,,
\end{eqnarray}
where the divisors are $D_i = (\tilde h_i),~(i=0,1,\cdots, n-1)$. The first term above recovers the original residues associated with $\Omega$ while the second term contains divisors generated solely by homogeneous polynomials. These divisors intersect at only an isolated point that is the origin. Conjecture \ref{conj:1} now applies to the second term straightaway.

According to our conjecture, the following differential operator $\mathbb D$ fully characterizes the relevant local information of the residue
\begin{align}
\mathbb D = \sum_{ \{s_i\} } a_{s_0\cdots s_{n-1}} \partial^{s_0}_{0}\cdots \partial^{s_{n-1}}_{n-1} \,,
\end{align}
where the summation is taken over all the non-negative solutions to the Frobenius equation $s_0+\cdots s_{n-1}=\text{ord}(\mathbb D)$ and
\begin{align}
\text{ord}(\mathbb D) = \text{deg}(\tilde h_1)+\cdots+\text{deg}(\tilde h_{n-1})+\text{deg}(\tilde h_n)-n\,.
\end{align}
The coefficients $a_{s_0 \cdots s_{n-1}}$ in the operator $\mathbb D$ are uniquely fixed by the local duality theorem and the intersection number of the divisors $D_i$'s. 

The local duality theorem yields,
\begin{eqnarray}
 0&=&\oint\limits_{{\tilde h}_0={\tilde h}_1=\cdots= {\tilde h}_{n-1}=0} \frac{P\, Q \,d\sigma_0\wedge d\sigma_1\wedge\cdots \wedge d\sigma_{n-1}}{ {\tilde h}_0\cdots {\tilde h}_{n-1}  } \,, \label{localdualityth}
\end{eqnarray}
where $Q$ is a polynomial of degree $d_Q$ in the ideal $\langle {\tilde h}_0, {\tilde h}_1,\cdots, {\tilde h}_{n-1}\rangle$ and $P$ is a holomorphic function in the neighbourhood of the intersecting point. 
To extract enough constraints from (\ref{localdualityth}), we need at most have $Q$ run over the homogeneous polynomials $\tilde h_i$ and $P$ all the monomials of degree $M-d_Q$ for each $Q$. The residue of $PQ$ vanishes in each case. Namely the constraints read,
\begin{eqnarray}
\mathbb{D} \Big( {\tilde h}_j\prod_{i=0}^{n-1}\sigma_i^{r_i} \Big)=0\,, \quad j=0,\cdots, n-1\,, \label{homoEqn}
\end{eqnarray}
for all possible solutions to the equation $\sum_{i=0}^{n-1} r_i =M-\text{deg}(\tilde h_j)$ where all $r_i$'s are non-negative integers. These constraints are not completely independent, in fact they have a large redundancy. By probing a large number of non-trivial examples, we observed that these equations are enough to fix the coefficients $a_{s_0\cdots s_{n-1}}$ up to a global scalar factor.
\footnote{We have not yet found a simple way to show that the rank of the homogeneous constraints is exactly $(\text{the number of } a_{s_0\cdots s_{n-1}}) -1$. We have observed this property in all the examples considered and believe this is true in general. In principle, if the homogeneous constraints are constructed with the generators of $\mathcal O/I$ ($\mathcal O$ is the polynomial ring of the ideal $I=\langle \tilde h_0, \cdots, \tilde h_{n-1}\rangle$) , these constraints are of less redundancy. This property may help in determining the rank of the homogenous constraints.
}

We only need one more inhomogeneous equation to determine this factor and the intersection number defined as follows is a natural choice, 
\begin{eqnarray}
\prod_{i=0}^{n-1} \text{deg} (\tilde h_i)=\oint\limits_{{\tilde h}_0=\cdots={\tilde h}_{n-1}=0} {d{\tilde h}_0\wedge \cdots \wedge d{\tilde h}_{n-1}\over  {\tilde h}_0\cdots {\tilde h}_{n-1} }\,. 
\end{eqnarray}
This translates to the inhomogeneous constraint,
\begin{eqnarray}\label{inhomoEqn}
\mathbb{D} \left[\det({\partial_i {\tilde h}_j})\right]=\prod_{i=0}^{n-1} \text{deg} (\tilde h_i)\,.
\end{eqnarray}

Combining the independent equations from (\ref{homoEqn}) and (\ref{inhomoEqn}), we have just enough constraints to determine the coefficients $a_{s_0\cdots s_{n-1}}$. Therefore we have obtained the value of the residue at the origin, i.e
\begin{eqnarray}
\Res_{ \{D_1,\cdots, D_{n-1}, D_n \},p'} ~\tilde\Omega  = \mathbb D \left[ \frac{\mathcal N_{reg}(\sigma_i)}{\sigma_0 -1} \right]\,.
\end{eqnarray}

\subsection{Residues at infinity}
Now we move on to the second term in (\ref{rewritten1loopAwithh}). Due to the degree of $\mathcal N_\infty$, we need to deal with the poles at infinity.

According to the global residue theorem, the residues of a meromorphic $n$-form on a compact $n$-dimensional complex manifold sum up to zero. This gives an equation relating residues at different poles (of the form) on the manifold, and thus provides an alternative method to compute residues: if we are interested in the sum of residues at certain poles, we can compute that of all the other poles on the manifold instead. 

Since we are dealing with forms on $\mathbb{C}^n$ which is non-compact, to use the method above, we embed $\mathbb{C}^n$ into a compact n-dim manifold and use the global residue theorem there. A natural choice for the compact manifold is $\mathbb{CP}^n$, where $\mathbb{C}^n$ can be identified with one of the standard coordinate patches, say $U_0=\{[z_i]_{i=0,\ldots,n}\in \mathbb{CP}^n|z_0\neq 0\}$, of $\mathbb{CP}^n$. Then the point(s) at infinity are simply those in the complement of $U_0$, i.e $\{[z_i]_{i=0,\ldots,n}\in \mathbb{CP}^n|z_0=0\}$. In order to use the global residue theorem, we should not only extend the manifold from $\mathbb{C}^n$ to $\mathbb{CP}^n$, but also extend the original differential form to the whole $\mathbb{CP}^n$. Namely, we now regard the original differential form as a \emph{local} expression on the coordinate patch $U_0$, and extend\footnote{which will become clear in later example.} it naturally to $\mathbb{CP}^n$ by the homogenous coordinates. Depending on the original differential form, the extended form may develop poles at infinity, and in that case the global residue theorem simply says that the sum of residues of finite poles and that of poles at infinity is zero. Note that now those points at infinity is in fact no different from those in $\mathbb{C}^n$, they are at infinity only w.r.t the patch $U_0$. In summary, to compute 
the residue of a pole at infinity we  need go to a patch
 covering that  point and then compute as usual.

However, the discussion above is usually not convenient in practice. Here we introduce another method. 
Suppose we want to compute the sum of residues of all finite 
poles of the differential form,
\begin{align}
\Omega=\frac{1}{h_1\ldots h_{n-1} } \frac{\mathcal N_\infty}{\mathcal D} d\sigma_1\wedge\ldots\wedge d\sigma_{n-1} \,.
\end{align}
Then we can consider the global residue theorem for the form
\begin{align}
\Gamma=\frac{\mathcal N_\infty}{\tilde{h}_1 \ldots \tilde{h}_{n-1} \tilde h_0}d\sigma_1\wedge\ldots\wedge d\sigma_{n-1}\wedge d\sigma_0~,
\end{align}
where $h_0:=\tilde{\mathcal D}\sigma_0^m(\sigma_0-1)$, $\tilde{h}_i$ and $\tilde{\mathcal D}$
 mean the homogenized version of $h_i$ and $\mathcal D$ as 
 in~(\ref{eq:Homgen}) and $m$ is a positive integer such that the following equation is valid
\begin{align}
\text{deg}(\mathcal N_\infty)=\text{deg}(\tilde h_1)+\text{deg}(\tilde h_2)+\ldots+\text{deg}(\tilde h_{n-1})+\text{deg}(h_0)-(n+1).
\end{align}
With these choices, the form $\Gamma$ has no pole at 
infinity and the sum of all finite residues simply vanishes.
 Now  the set of poles of $\Gamma$ consists of two parts
\begin{align}
&S_0=\{\tilde{h}_1=\ldots=\tilde{h}_{n-1}=0,\tilde{\mathcal D}\sigma_0^m=0\}\\
&S_1=\{\tilde{h}_1=\ldots=\tilde{h}_{n-1}=0,\sigma_0=1\}.
\end{align}
The global residue theorem thus reads
\begin{align}
\sum_{p\in S_0}\text{Res}_{\{(\tilde{h}_1),\ldots,(\tilde{h}_{n-1}), (\sigma_0^m)\},p}\Gamma+
\sum_{p'\in S_1}\text{Res}_{\{(\tilde{h}_1),\ldots,(\tilde{h}_{n-1}), (\sigma_0-1)\},p'}\Gamma=0
\end{align}
The second term actually equals to the sum of residues of 
all the finite poles of $\Omega$, which can been seen by 
writing it in terms of contour integral and then integrate 
out $\sigma_0$:
\begin{align}
\sum_{p'\in S_1}\text{Res}_{\{(\tilde{h}_1),\ldots,(\tilde{h}_{n-1}), (\sigma_0-1)\},p'}\Gamma&=
\oint\limits_{\tilde{h}_1=\ldots=\tilde{h}_{n-1}=0,\sigma_0=1}\frac{\mathcal N_\infty}{\tilde{h}_1\ldots \tilde{h}_{n-1}\cdot h_0}d\sigma_1\wedge\ldots\wedge d\sigma_{n-1}\wedge d\sigma_0\nonumber\\
&=\oint\limits_{h_1=\ldots=h_{n-1}=0}\frac{1}{h_1\ldots h_{n-1}} {\mathcal N_\infty \over \mathcal D} d\sigma_1\wedge\ldots\wedge d\sigma_{n-1}\nonumber\\
&=\sum_{\text{finite poles}}\text{Res}_{\{h_1,\ldots,h_{n-1}\},p}~\Omega~.
\end{align}
The first term, on the other hand, is of the form meeting conditions of Conjecture~\ref{conj:1} and therefore can be computed accordingly. 
Thus the problem of finding the sum of residues of finite 
poles of $\Omega$ has been turned into that 
for $\Gamma$ which can be done using Conjecture~\ref{conj:1}. 

\subsection{Residues at spurious poles}\label{sec:residue_at_spurious_poles}
Now we are left with the last residue at the pole $\sigma_1=\cdots\sigma_{n-1}=1$. We call this pole ``spurious" since it is not present in the solution to the original scattering equations. 
Notice that here we are interested in the residue \emph{at a particular pole}, and thus the previous ansatz (\ref{defD}) does not apply because it computes the sum of residues of \emph{all} finite poles. To deal with the current case, first we need to consider another homogenized version of the polynomial scattering equations $\{\hat h_i\}$. From $\{\hat{h}_i\}$ we then construct a differential operator $\hat{\mathbb{D}}$.\footnote{This may sound the same with the process dealing with the finite poles, i.e $\{h_i\}\rightarrow \{\tilde h_i\}\rightarrow \mathbb D$, but actually both $\{\hat h_i\}$ and $\hat{\mathbb{D}}$ are constructed in ways different from the previous versions, as explained in detail in the following.} This operator is conjectured to give the residue at the spurious pole.

As we have mentioned above, the residue at a given pole depends only on the local information of that pole. Hence a natural step to take is to parallel transport the coordinate system by $\sigma_i \rightarrow \sigma_i +1$ such that the pole becomes the origin. This is convenient for later discussion in this part.

Here we construct $\{\hat{h}_i\}$ from $\{h_i\}$. In the neighbourhood of the pole, now at the origin, each polynomial $h_i$ can be separated into its leading order term $\text{L}(h_i)$ and higher order terms $\text{H}(h_i)$. 
Let $d_{L_i}$ be the degree of the leading order term $\text{L}(h_i)$. For each term $m$ in $\text{H}(h_i)$ of degree $d_m$, we construct $\hat m$ by substituting some of the $\sigma_i$ factors with constants (w.r.t $\sigma$'s) $t_i$'s such that $\hat m$ has a power in $\sigma$'s the same with $L(h_i)$. \footnote{For instance, if $m=\sigma_1\sigma_2 \sigma_3^2$ and $d_{L_i}=2$, the replacement can be $\hat m=t_1 t_2 \sigma_3^2$. Or replace $\sigma_2\cdot\sigma_2$ with $t_1\sigma_2$ if $d_{L_i}=1$. Note that the particular choice of the replaced variables does not affect the final result. We have yet not found a direct proof for this property, but have checked it against many numerical or analytical examples.}
In this way all higher order terms are degraded to the degree of the leading term, with which we then define the homogenous polynomials $\{\hat h_i\}$
\begin{eqnarray}
\text{H} (h_i) = \sum_a m_a(\sigma_i) \rightarrow \hat{\text{H}}(h_i) =\sum_a \hat m_a(\sigma_i,t_i)\,, \quad\quad\quad \hat h_i = \text{L}(h_i) + \hat{\text{H}}(h_i)\,. \label{hatdef}
\end{eqnarray}

Now we proceed to define the operator $\hat{\mathbb{D}}$. Those hatted polynomials $\hat h_i$'s are homogeneous, so exactly as in Section (\ref{sec:operator}) we could define the differential operator ${\mathbb{D}}$ corresponding to the integration with $\hat h_i$'s. ${\mathbb{D}}$ is a differential operator with coefficients $a_{s_0\cdots s_{n-1}}$ being functions of $t_i$'s. Namely,
\begin{eqnarray}
\oint\limits_{\hat h_1=\cdots =\hat h_{n-1}=0 } \frac{d\sigma_1\wedge \cdots \wedge d\sigma_{n-1}}{\hat h_1 \cdots \hat h_{n-1}}\;\Longleftrightarrow \; \mathbb{D}=\sum_{ \{s_i \} }a_{s_1\cdots s_{n-1}}(t_j)\partial^{s_1}_{r_1}\partial^{s_2}_{r_2}\cdots \partial^{s_{n-1}}_{r_{n-1}}\,.
\end{eqnarray}
From ${\mathbb{D}}$ we now define the differential operator $\mathbb{\hat D}$ that acts on an arbitrary function $\mathcal F$ in the following way,
\begin{eqnarray}
\mathbb{\hat D} \mathcal F = \sum_{ \{ s_0,\cdots s_{n-1} \} }\partial^{s_1}_{r_1}\partial^{s_2}_{r_2}\cdots \partial^{s_{n-1}}_{r_{n-1}}\left( a_{s_1\cdots s_{n-1}}(\sigma_j) \mathcal F \right)\, .
\end{eqnarray}

Based on large amount of numerical test, we propose the following conjecture to compute the residue at a particular pole:
\begin{conjecture}\label{conj:2}If the solution of $\langle L(h_0)=0, L(h_1)=0,\cdots, L(h_{n-1})=0\rangle$ is an isolated point $p$, then the residue at this pole defined by
\begin{eqnarray}
\Res_p[\mathcal{R}(\sigma_i)]:=\oint\limits_{{ h}_1=\cdots={h}_{n-1}=0} {d\sigma_1 \wedge \cdots \wedge d\sigma_{n-1}\over  { h}_1\cdots {h}_{n-1} } \mathcal{R}(\sigma_i)
\end{eqnarray}
can be obtained by $\hat{\mathbb{D}}$ in the following way
\begin{eqnarray}
\Res_p[\mathcal{R}(\sigma_i)]= \mathbb{\hat D} \left[\mathcal{R}(\sigma_i)\right]\,
\label{Ansatz2}
\end{eqnarray}
where $\mathcal R(\sigma_i)$ is a holomorphic function in the neighborhood of the origin.
\end{conjecture}
Besides a large number of numerical checks, this conjecture has also passed the analytical check for the one-loop super Yang-Mills integrand with four external particles.
\footnote{It is possible to get some intuition about this conjecture by considering a simpler scenario. Let $f_i$ be a set of inhomogeneous polynomials and $\text{L} (f_i)$ their respective leading order terms. The polynomials $\hat f_i$ are defined the same way as in (\ref{hatdef}). Suppose there exists a transformation matrix $\mathbb M$ such that,
\begin{eqnarray}
\left(
\begin{array}{c}
\vdots\\
\text{L} (f_i) \\
\vdots
\end{array}
\right) = \mathbb M (\sigma_i) \left(
\begin{array}{c}
\vdots\\
f_i \\
\vdots
\end{array}
\right) \,, \quad\quad 
\left(
\begin{array}{c}
\vdots\\
\text{L} (f_i) \\
\vdots
\end{array}
\right) = \mathbb M (t_i) \left(
\begin{array}{c}
\vdots\\
\hat f_i \\
\vdots
\end{array}
\right)\,. \nonumber
\end{eqnarray}
Then we have, for a rational function $\mathcal F$
\begin{eqnarray}
&& \oint\limits_{\cdots\cap (f_i)\cap\cdots} \frac{d\sigma_1\cdots d\sigma_i\cdots}{f_1\cdots f_i\cdots} \mathcal F =
\oint\limits_{\cdots\cap (\text{L}(f_i)) \cap\cdots} \frac{\det\mathbb M(\sigma_i)\, d\sigma_1\cdots d\sigma_i\cdots}{\text{L}(f_1)\cdots \text{L}(f_i)\cdots}\mathcal F \,, \nonumber\\
&& \oint\limits_{\cdots\cap (\hat f_i)\cap\cdots} \frac{d\sigma_1\cdots d\sigma_i\cdots}{\hat f_1\cdots \hat f_i\cdots} \mathcal F =
\oint\limits_{\cdots\cap (\text{L}(f_i)) \cap\cdots} \frac{\det\mathbb M(t_i)\, d\sigma_1\cdots d\sigma_i\cdots}{\text{L}(f_1)\cdots \text{L}(f_i)\cdots}\mathcal F \,. \nonumber
\end{eqnarray}
Let $\mathbb D$, $\mathbb D_L$ and $\mathbb{\hat D}$ denote the differential operators corresponding to the integrations with $f_i$, $\text{L}(f_i)$ and $\hat f_i$ in the denominator respectively. It can be verified that,
\begin{eqnarray}
\mathbb D \mathcal F= \mathbb D_L \left(\det\mathbb M(\sigma_i) \mathcal F\right) = \mathbb D_L \left( \lim_{t_i\rightarrow \sigma_i}\det \mathbb M (t_i) \mathcal F \right)=\mathbb{\hat D} \left( \lim_{t_i\rightarrow \sigma_i}\det \mathbb M (t_i) \mathcal F \right)\,.\nonumber
\end{eqnarray}
Of course it is not obvious whether such a transformation exists and we can not prove our conjecture in general.
}

\section{Four-point one-loop SYM integrand}
\label{sec:fourpoint}
In this section, we further exemplify our algorithm by studying the super Yang-Mills one-loop amplitude for four particles. The four-point SYM amplitude is known to be particularly simple, since it only has one non-vanishing helicity configuration that is MHV. The planar 4-point amplitude at one loop is captured by the BDS ansatz \cite{Bern:2005iz} while the non-planar contribution is shown to be related to the planar part in \cite{Arkani-Hamed:2014via,Franco:2015rma,Chen:2015qna,Chen:2015bnt,Bourjaily:2016mnp}. 

Here we are interested in mainly the mathematical properties of the CHY-form of this one-loop integrand and its explicit expression is given in~\cite{Geyer:2015bja, Geyer:2015jch},
\begin{eqnarray}
\mathcal I_4^{l=1} =\oint\limits_{f_1=\cdots=f_{3}=0} {d\sigma_1\cdots d\sigma_{3} \over  f_1\cdots f_{3}} ~\text{Pf}(M_4)~PT_4 \prod_{i=1}^{3} {1\over \sigma_i} \,.
\end{eqnarray}
The well-known Parke-Taylor factor reads 
\begin{equation}
    PT_4=\sum_{\gamma}{1\over \sigma_{\gamma(1)} (\sigma_{\gamma(1)}-\sigma_{\gamma(2)})(\sigma_{\gamma(2)}-\sigma_{\gamma(3)})(\sigma_{\gamma(3)}-\sigma_{\gamma(4)})},
\end{equation}
where we sum over all the $S_4$ permutations of the indices.
The Pfaffian in this case is simply a constant. The polynomial scattering equations are easy to construct as in Section \ref{sec:polynomialSE}.
Substituting the Vandermonde determinant and the Parke-Taylor factor, we arrive at the simple contour integral form of the 4-point integrand,
\begin{eqnarray}\label{eq:A4contour}
\mathcal I^{l=1}_4 = \oint\limits_{h_1 = h_2 =h_3=0} \frac{d\sigma_1 \wedge d\sigma_2 \wedge d\sigma_3}{h_1 h_2 h_3}\, \mathcal R_4 -\oint\limits_{\sigma_1=\sigma_2=\sigma_3=1} \frac{d\sigma_1 \wedge d\sigma_2 \wedge d\sigma_3}{h_1 h_2 h_3} \,\mathcal R_4  \,,
\end{eqnarray}
where the polynomials $h_i$'s are given in (\ref{eqn:hdef}) and we have chosen the gauge $\sigma_4=1$. The function $\mathcal R_4$ reads,
\begin{eqnarray}
\mathcal R_4 =
\frac{(\sigma_2-1) (\sigma_1-\sigma_3) \left[\sigma_1^2 \sigma_3 + \sigma_2 \sigma_3 + \sigma_1 \sigma_2 \left(\sigma_2+\sigma_3 (\sigma_3-4) \right) \right]}{\sigma_1 \sigma_2 \sigma_3} \,.
\end{eqnarray}

We consider the integral around the origin first. The integrand is a sum of meromorphic functions and the terms can be put into three categories, depending on their singularities: (1) functions that have only poles originating from the scattering equations, i.e poles of $h_i$'s; (2) functions that have poles originating from $h_i$'s and other factors, such as $\sigma_i$'s, but are regular at infinity; (3) functions that have poles at infinity. The residues of those in the first category are obviously zero and we drop these terms from now on.
There are only four surviving terms and we denote them as $\mathcal R_4 =\sum^{4}_{i=1}\mathcal R_4^{(i)}$. The first three terms read,
\begin{eqnarray}
\mathcal R^{(1)}_4 = \frac{(\sigma_2 -1) \sigma_3}{\sigma_1} \,, \quad\quad
\mathcal R^{(2)}_4 = \frac{(\sigma_1-\sigma_3) \sigma_1}{\sigma_2}\,, \quad\quad
\mathcal R^{(3)}_4 = \frac{\sigma_1\sigma_2 (\sigma_2-1)}{\sigma_3} \,.
\end{eqnarray}
These terms do not have non-zero residues at infinity. The residue corresponding to the last one, however, is non-vanishing at infinity and needs to be taken care of differently,
 \begin{eqnarray}
 \mathcal R_4^{(4)} = \sigma_2 \sigma_3 (\sigma_3 -\sigma_1)\,.
 \end{eqnarray}
 The meromorphic forms corresponding to $\mathcal{R}_4^{(i)}$ are
 \begin{align}
\Gamma_1&=\frac{(\sigma_2 -1) \sigma_3}{h_1 h_2 h_3\cdot\sigma_1} d\sigma_1\wedge d\sigma_2 \wedge d\sigma_3,\\
\Gamma_2&=\frac{(\sigma_1-\sigma_3) \sigma_1}{h_1 h_2 h_3\cdot\sigma_2} d\sigma_1\wedge d\sigma_2 \wedge d\sigma_3\\
\Gamma_3&=\frac{\sigma_1\sigma_2 (\sigma_2-1)}{h_1 h_2 h_3\cdot\sigma_3} d\sigma_1\wedge d\sigma_2 \wedge d\sigma_3,\\
\Gamma_4&=\frac{\sigma_2 \sigma_3 (\sigma_3 -\sigma_1)}{h_1 h_2 h_3} d\sigma_1\wedge d\sigma_2 \wedge d\sigma_3,
\end{align}
 
 \subsection{Computing the residues at finite poles}
 \label{FinitTerm}
The residues associated with $\Gamma_1,~\Gamma_2$, and $\Gamma_3$ can all be obtained the same way, following our conjecture, and here we just take $\Gamma_1$ as an example.
First, we homogenize the factors $h_1,h_2,h_3$ with an auxiliary variable $\sigma_0$ and the residue associated with $\Gamma_1$ becomes,
\begin{eqnarray}
\oint\limits_{h_1 = h_2 =h_3 =0} \Gamma_1 = \oint\limits_{\tilde h_1 =\tilde h_2 =\tilde h_3 =\sigma_0-1} \tilde\Gamma_1
\end{eqnarray}
where
\begin{eqnarray}
\tilde\Gamma_1 = \frac{(\sigma_2 -1) \sigma_3}{\tilde h_1 \tilde h_2 \tilde h_3\cdot\sigma_1 (\sigma_0-1)}  d\sigma_1\wedge d\sigma_2 \wedge d\sigma_3\wedge d\sigma_0\,.
\end{eqnarray}
The global residue theorem yields,
\begin{eqnarray}
\oint\limits_{\tilde h_1 =\tilde h_2 =\tilde h_3 =\sigma_0-1} \tilde\Gamma_1 = -\Res_{ \{D_1, D_2, D_3, (\sigma_0) \},p }~\tilde\Gamma_1\,.
\end{eqnarray}
where $D_i = (\tilde h_i)$ ($i=1,2,3$). All the intersecting divisors on the right-hand-side are generated by homogeneous polynomials whose common zero is assumed to be a single point, which thus must be the origin. The right-hand-side is corresponding to a third-order differential operator, as conjectured in \ref{conj:1},
\begin{eqnarray} 
\mathbb D = \sum_{\updown{0\leq s_i\leq 3\,,} {\;s_0+s_1+s_2+s_3=3}} a_{s_0 s_1 s_2 s_3}\frac{\partial^{s_0}}{\partial\sigma_0^{s_0}} \frac{\partial^{s_1}}{\partial\sigma_1^{s_1}} \frac{\partial^{s_2}}{\partial\sigma_2^{s_2}} \frac{\partial^{s_3}}{\partial\sigma_3^{s_3}} \,.
\end{eqnarray}
There are 20 coefficients $a_{s_0 s_1 s_2 s_3}$ to be determined. The local duality theorem yields,
\begin{eqnarray}
\mathbb D \left( \sigma_i \sigma_j \tilde h_0 \right) = \mathbb D \left(\sigma_i \sigma_j \tilde h_1\right) = \mathbb D \left(\sigma_i \tilde h_2 \right) = \mathbb D \; \tilde h_3 =0\,\quad\quad 0\leq i,j \leq 3\,.
\end{eqnarray}
These constraints have a huge redundancy and if one carefully computes the rank of the constraint matrix, it is in fact 19. The intersection number of the divisors in this case is 6 and this demands,
\begin{eqnarray}
\mathbb D \left( \det\left[\frac{\partial\tilde h_i}{\partial\sigma_j}\right]\right) = 6 \,.
\end{eqnarray}
Now we have 20 conditions that fix the coefficients completely. These constraints are simple and linear conditions and solving them possesses no difficulty at all. Substituting the solution into $\mathbb D$ the residue associated with $\Gamma_1$ is given by
\begin{eqnarray}
 -\Res_{ \{D_1, D_2, D_3, (\sigma_0) \},p }~\tilde\Gamma_1&=& -\mathbb D \left[ \frac{(\sigma_2-1)\sigma_3}{\sigma_0 -1}\right] = -\mathbb D \left[ \frac{-\sigma_3}{\sigma_0 -1}\right] \nonumber\\
&=& \frac{1}{\ell\cdot k_1 \, \ell\cdot k_4 \,(k_{12} + \ell\cdot k_1 +\ell\cdot k_2)} \, .
\end{eqnarray}
Likewise, the terms $\mathcal R^{(2)}_4$ and $\mathcal R^{(3)}_4$ give rise to the following residues respectively,
\begin{eqnarray}
&& -\mathbb D \left[ \frac{\sigma_1 (\sigma_1 -\sigma_3)}{\sigma_0 -1} \right] =-\mathbb D \left[ \frac{\sigma^2_1}{\sigma_0 -1} \right]= \frac{1}{\ell\cdot k_1\, \ell\cdot k_2 \, (k_{23} + \ell\cdot k_2 +\ell\cdot k_3)} \,, \\
&&- \mathbb D \left[ \frac{\sigma_1\sigma_2 (\sigma_2-1)}{\sigma_0-1} \right] = - \mathbb D \left[ \frac{\sigma_1\sigma^2_2 }{\sigma_0-1} \right] =\frac{1}{\ell\cdot k_2\, \ell\cdot k_3 ( k_{34} + \ell\cdot k_3 +\ell\cdot k_4)} \,.
\end{eqnarray}
 
\subsection{Computing the residues at infinity}
Now we discuss the term $\mathcal R_4^{(4)}$ which has a 
non-zero residue at infinity. The standard method to obtain  
the residue at infinity is discussed in Appendix~\ref{STInfity}. Here we obtain the residue also by our ansatz.  For that, we consider the form
\begin{align}
\tilde\Gamma_4=\frac{\left(\sigma_3-\sigma_1\right) \sigma_2 \sigma_3}{\tilde{h}_1 \tilde{h}_2 \tilde{h}_3\cdot\sigma_0(\sigma_0-1)} d\sigma_1\wedge d\sigma_2 \wedge d\sigma_3\wedge d\sigma_0.
\end{align}
According to the global residue theorem, we know that its residue at infinity is zero, i.e residues of all finite poles sum up to zero (\cite{hartshorne2013algebraic, griffiths2014principles}). We choose the four divisors (for this 4-dim space) by
\begin{align}
D_1=(\tilde{h}_1),~
D_2=(\tilde{h}_2),~
D_3=(\tilde{h}_3),~
D_4=(\sigma_0 (\sigma_0-1)).
\end{align}
The global residue theorem leads to
\begin{eqnarray}\label{eq: GRT_2edMethod}
0 &=&\sum_{p}\text{Res}_{\{D_1,D_2,D_3, (\sigma_0(\sigma_0-1) ) \},p}~\tilde\Gamma_4 \nonumber\\
&=&
\sum_{p}\text{Res}_{\{D_1,D_2,D_3, (\sigma_0) \},p}~\tilde\Gamma_4+\sum_{p'}\text{Res}_{\{D_1,D_2,D_3, (\sigma_0-1) \},p'}~\tilde\Gamma_4 \,.
\end{eqnarray}
The second term, if integrated over $\sigma_0$ first, simply returns to the original integral 
$$\oint\limits_{h_1 = h_2 =h_3=0} \frac{d\sigma_1 \wedge d\sigma_2 \wedge d\sigma_3}{h_1 h_2 h_3}\, \mathcal R^{(4)}_4$$
Now according the anstaz \ref{defD}, the first term is explicitly
\begin{align}\label{eq: res_result_method2}
 \mathbb D_3 \left[ \frac{\sigma_2 \sigma_3(\sigma_3 -\sigma_1)}{\sigma_0 -1} \right] =\mathbb D_3 \left[ \frac{\sigma_2 \sigma^2_3}{\sigma_0 -1} \right]= -\frac{1}{l\cdot k_3\left( -k_{23}+l\cdot k_2+l\cdot k_3 \right)\left( -l\cdot k_4 \right)}.
\end{align}
Thus from (\ref{eq: GRT_2edMethod}) we immediately get
$$\oint\limits_{h_1 = h_2 =h_3=0} \frac{d\sigma_1 \wedge d\sigma_2 \wedge d\sigma_3}{h_1 h_2 h_3}\, \mathcal R^{(4)}_4=\frac{1}{l\cdot k_3\left( -k_{23}+l\cdot k_2+l\cdot k_3 \right)\left( -l\cdot k_4 \right)}$$

\subsection{Computing the residues at spurious poles}
We are now left with the spurious pole at $\sigma_1=\cdots=\sigma_{n-1}=1$. The parameter transformation $\sigma_i\rightarrow \sigma_{i}+1$ shifts the pole to the origin. The polynomial scattering equations are directly read off from (\ref{eq:polynomialScatEq}).
Unlike the first integral in (\ref{eq:A4contour}), now we have to take all the terms in $\mathcal R_4$ into consideration. That is, we are computing the residue at the origin associate with the form,
\begin{eqnarray}
\Gamma_{spurious} = \frac{d\sigma_1 \wedge d\sigma_2 \wedge d\sigma_3}{h_1(\sigma_i+1) h_2(\sigma_i+1) h_3(\sigma_i+1)} \,\mathcal R_4(\sigma_i+1)\,.
\end{eqnarray}
Following the discussion in Section \ref{sec:residue_at_spurious_poles}, we homogenize the shifted polynomials $h(\sigma_i+1)$ as follows, 
\begin{eqnarray}
&&\hat h_1=l_1\sigma_1 +l_2\sigma_2 +l_3\sigma_3 \, ,\nonumber\\
&&\hat h_2=-l_{12} \sigma_1 \sigma_2 -l_{13} \sigma_1 \sigma_3 -l_{23}\sigma_2 \sigma_3 \, ,\nonumber\\
&&\hat h_3=- l_4 t_1\sigma_2 \sigma_3 +k_{12} \sigma_1 \sigma_2+k_{13} \sigma_1\sigma_3+k_{23} \sigma_2 \sigma_3\, . 
\end{eqnarray}
At the moment the quantity $t_1$ in $\hat{h}_3$ is regarded as a number and the degrees of the polynomials are $\text{deg}(\hat h_1)=1$, $\text{deg}(\hat h_2)=2$ and $\text{deg}(\hat h_3) =2$. The intersection number of the divisors $\hat D_i=(\hat h_i)\, , ~(i=1,2,3)$ is $4$. Now we consider the following residue,
\begin{eqnarray}
\Res_{ \{ \hat D_1, \hat D_2, \hat D_3\}, p} \hat\Gamma_{spurious} \,, \quad\quad \text{with}~~~
\hat\Gamma_{spurious} = \frac{d\sigma_1 \wedge d\sigma_2 \wedge d\sigma_3}{\hat h_1 \hat h_2 \hat h_3} \mathcal R_4 (\sigma_i+1)\,.
\end{eqnarray}
To obtain this residue using Conjecture \ref{conj:2}, a second-order differential operator is to be worked out,
\begin{eqnarray}
\mathbb{D}=a_{002}{\partial^2\over\partial\sigma_3^2}+a_{011}{\partial^2\over\partial\sigma_2\partial\sigma_3}+a_{020}{\partial^2\over\partial\sigma_2^2}+a_{101}{\partial^2\over\partial\sigma_1\partial\sigma_3}+a_{110}{\partial^2\over\partial\sigma_1\partial\sigma_2}+a_{200}{\partial^2\over\partial\sigma_1^2}\, .
\end{eqnarray}
The local residue theorem and the intersection number conditions demand,\begin{equation}
\mathbb{D}  (\sigma_1 \hat h_1)=0\, , ~\mathbb{D} (\sigma_2 \hat h_1)=0\, ,~\mathbb{D}( \sigma_3 \hat h_1)=0\,,~\mathbb{D}  \hat h_2=0\,,~\mathbb{D}  \hat h_3=0\,, ~\mathbb{D} \hat{\mathcal J}=4\,.
\end{equation}
where $\hat{\mathcal J}\equiv \det(\partial_i  \hat h_j).$
These constraints are easily solved. Note that the condition matrix here is invertible as $t_1\rightarrow0$. Substituting $t_1=\sigma_1$ back in the expressions for the coefficients $a_{ijk}$'s, the ansatz (\ref{Ansatz2}) for the inhomogenous case leads to,
\begin{eqnarray}
 && \hat{\mathbb D}\, (\mathcal I_4(\sigma_i+1) )\nonumber\\
&\equiv& {\partial^2\over\partial\sigma_3^2}\left[a_{002}\mathcal R_4(\sigma_i+1)\right]+{\partial^2\over\partial\sigma_2\partial\sigma_3}\left[a_{011}\mathcal R_4(\sigma_i+1)\right]+{\partial^2\over\partial\sigma_2^2}\left[a_{020}\mathcal R_4(\sigma_i+1)\right]\nonumber\\
&& +{\partial^2\over\partial\sigma_1\partial\sigma_3}\left[a_{101}\mathcal R_4(\sigma_i+1)\right]+{\partial^2\over\partial\sigma_1\partial\sigma_2}\left[a_{110}\mathcal R_4(\sigma_i+1)\right]+{\partial^2\over\partial\sigma_1^2} \left[a_{200}\mathcal R_4(\sigma_i+1)\right] \nonumber\\
&=&\Res_{\{ (h_1(\sigma_i+1)), (h_2(\sigma_i+1)), (h_3(\sigma_i+1)) \} ,p} \Gamma_{spurious} 
\end{eqnarray}
%
Fortunately, the explicit expressions for the coefficients $a_{ijk}$ are not necessary for computing this residue. For instance, consider the term
\begin{eqnarray}
 {\partial^2\over\partial\sigma_1\partial\sigma_3}a_{101}\mathcal R_4(\sigma_i+1)|_{\sigma_i=0}
=\left[\frac{1}{\sigma_2+1}+\frac{1}{(\sigma_3+1)^2}+\frac{1}{\sigma_1+1}-3\right] \left.{\partial a_{101}(\sigma_1)\over \partial\sigma_1}\right|_{\sigma_i=0}.
\end{eqnarray}
This vanishes since ${\partial a_{101}(\sigma_1)\over \partial\sigma_1}$ is holomorphic in the neighbourhood of the origin and the factor multiplying this derivative is zero at the origin. This is true for all the terms in the action of $\hat{\mathbb D}$ when the condition matrix is invertible. Hence the residue at the spurious pole is vanishing.

\subsection{Summary of the four-point integrand}
Now we conclude this section by summarizing the CHY-form for the 4-point 1-loop SYM integrand evaluated by our ansatz. Explicitly, the final result reads,
\begin{eqnarray}
\mathcal I_4^{l=1} &=&\oint\limits_{h_1=h_2=h_3=0}\left(\Gamma_1+\Gamma_2+\Gamma_3+\Gamma_4 \right) \nonumber\\
&=&~\frac{1}{\ell\cdot k_{1} \ell\cdot k_{4} (\ell\cdot k_{1}+\ell\cdot k_{2}+k_{12})}+\frac{1}{\ell\cdot k_{1} \ell\cdot k_{2} (\ell\cdot k_{2}+\ell\cdot k_{3}+k_{23})}\nonumber\\
&&-\frac{1}{\ell\cdot k_{2} \ell\cdot k_{3} (\ell\cdot k_{1}+\ell\cdot k_{2}-k_{12})}-\frac{1}{\ell\cdot k_{3} \ell\cdot k_{4} (\ell\cdot k_{2}+\ell\cdot k_{3}-k_{23})}\,.
\end{eqnarray}
The four terms in this expression have a one-to-one correspondence with the four forward-limit channels in the Q-cut representation of the same amplitude ~\cite{Baadsgaard:2015twa,Huang:2015cwh}. This is a consequence of the string origin of the CHY representations. The singular behaviour of the CHY-form is inherited from the worldsheet factorization structure~\cite{Berkovits:2013xba,Mason:2013sva,Geyer:2016wjx}, which is naturally related to the forward limit. 

\section{Outlooks}
So far we have developed a differential operator for the residue with respect to a general meromorphic form and exploited it in the study of the four-point CHY expressions at one loop. An immediate follow-up direction is to probe the one-loop CHY-form for higher points. Starting from 5-points, the integrands of amplitudes grow more complicated, in particular, nontrivial Pfaffians enter in the integrand in SYM and SUGRA.  Nevertheless, these factors are rational functions and our method is expected to apply to higher-point integrands comfortably. Having collected more analytical data for higher-point expressions, other ways may be discovered to determine the exact form of the differential operator, without solving the corresponding constraints by brute force.

Our method also serves as a useful tool to investigate the higher-loop CHY-forms in Yang-Mills and gravity theory, as well as to explore the non-planar regime of these theories where new symmetries are likely to be hiding. At two loops, the construction of the integral basis, involving the integration-by-parts (IBP) relations among loop integrands, remains an interesting open question. CHY-forms may be a new playground for such construction and the conjectures for residues hopefully help us in finding similar relations at the level of CHY expressions.
Many aspects of Yang-Mills and gravity outside the large-N limit are still uncharted at the moment. Constructing CHY-like representations for non-planar amplitudes is certainly of importance while the string origin of such representations may make some symmetries and algebraic structures, which are otherwise hard to observe, manifest.

Furthermore, this method also finds natural applications in a variety of aspects of scattering amplitudes, such as the Grassmannian integral form and the generalized unitarity cut.

\appendix
\section{Numerical Checks for the Ansatz}
In this appendix we present numerical checks of the ansatz for the differential operator used in the computation of the residues. To check the ansatz, we consider different ideals of polynomials $\langle h_1,h_2,\cdots, h_n\rangle$ and obtain the positions of the intersection point by solving the corresponding algebraic equations numerically. For some given integrand $g(\sigma_i)$ we compute its residues at these intersection points, both directly and using our conjectures. In all the examples we have considered, the numerical results obtained from both methods match beautifully.

To compute the residue directly at a given intersection point, we have to consider the nature of this intersection point first. There are two types of isolated intersection points, the singular ones and the non-singular ones. For a non-singular isolated point $p\in (h_1)\cap(h_2)\cdots\cap(h_n)$, the residue is 
\begin{equation}\label{eq:ResNonsin}
\Res_{p} {g(\sigma_i)d\sigma_1\wedge \cdots \wedge d\sigma_n\over h_1\cdots h_n}=\left.{g\over \mathcal J}\right |_p,
\end{equation} 
where $\mathcal{J}|_p=\left.\det\left({\partial f_i\over \partial \sigma_j}\right)\right |_p$ is nonzero for the non-singular point.  For a singular isolated intersection point, we need to perform a deformation first. To guarantee that no information of the singularity gets lost, we need a semiuniversal deformation and a deformation generated by the Tjurina algebra \cite{ebeling2007functions,greuel2007introduction} is a nice candidate. For a singular isolated point defined by a set of generators of the ideal $\langle h_1,h_2\cdots,h_n\in \mathcal{O}_{\mathbb{C}^n,p}\rangle$, the Tjurina algebra is 
$$
Tj=\mathcal{O}^n_{\mathbb{C}^n,p}/\langle h_1 \vec{e}_1,\cdots, h_1\vec{e}_n,\cdots,h_n \vec{e}_1,\cdots, h_n\vec{e}_n, \partial_{\sigma_1} \vec{h},\cdots, \partial_{\sigma_n} \vec{h}\rangle,
$$
 where $\vec{h}$ is the n-column $(h_1, h_2,\cdots, h_n)$ and $\vec{e}_i$ is the unit $n$-column with its $i$-th element set to be 1. In this case, the algebra has a  finite number of generators $\vec{g}_1, \cdots, \vec{g}_{\tau}$, where $\tau$ denotes the total number of the generators known as the Tjurina number. The semiuniversal deformation reads, 
 $$\vec{F}(\sigma, t)=\vec{h}(\sigma) +\sum_{i=1}^{\tau}t_i \vec{g}_i.$$
 In principle one is supposed to perform such a deformation. However, in practice, we only have to use a sub-space in  the Tjurina algebra such that the singular point of degree $d$ is decoupled into $d$ separated intersection points $p^i$. This process is easy to implement numerically by simply choosing each parameter $t_i$ to be a very small number. After deforming the singular point, we sum up the residue over all the separated intersection points and obtain the residue at the original point,
 \begin{equation}
 \label{eq:ResDeform}
\Res_{p} {g(\sigma_i)d\sigma_1\cdots d\sigma_n\over h_1\cdots h_n}=\sum_{i=1}^d \left.{g\over \mathcal J}\right|_{p^i}\,.
\end{equation}
In numerical computations, the high precision of the results is guaranteed as long as the deformation parameters $t_i$ are sufficiently small. 
\subsection{Homogeneous ideals}
In this section we take 6 homogeneous ideals as our examples to test Conjecture \ref{conj:1}. (We have tested our ansatz against a lot more examples and believe our method to be quite robust.) The coefficients in these ideals are randomly generated integers. Half of the ideals contain 3 variables and the degrees of the polynomials in them range from 4 to 6. The rest examples are ideals consisting of polynomials of degree 2 and the number of variables ranges from 4 to 6. The integrand is chosen to be $g(\sigma_i) = \frac{\sigma_1 +1}{\sum_{i=1}^{n} \sigma_i+1}$ for all examples, where $n$ denotes the number of the variables in the ideal.\footnote{This function $g$ is chosen such that none of the derivatives $\partial_{i}$ ever acts trivially on the integrand.}
The ideals are given below and the residues computed using the two methods are listed in Table \ref{HomId}. 
\begin{eqnarray*}
\mathcal{I}_3^4&=&\langle7 \sigma _1^4+7 \sigma _2 \sigma _1^3+9 \sigma _3 \sigma _1^3+2 \sigma _2^2 \sigma _1^2+18 \sigma _3^2 \sigma _1^2+11 \sigma _2 \sigma _3 \sigma _1^2+17 \sigma _2^3 \sigma _1+18 \sigma _3^3 \sigma _1~~~~~~~~~~~~~~~~~~\\
&&+23 \sigma _2 \sigma _3^2 \sigma _1+14 \sigma _2^2 \sigma _3 \sigma _1+9 \sigma _2^4+16 \sigma _3^4+20 \sigma _2 \sigma _3^3+19 \sigma _2^2 \sigma _3^2+12 \sigma _2^3 \sigma _3,\\
&&\sigma _1^4+23 \sigma _2 \sigma _1^3+12 \sigma _3 \sigma _1^3+13 \sigma _2^2 \sigma _1^2+22 \sigma _3^2 \sigma _1^2+22 \sigma _2 \sigma _3 \sigma _1^2+6 \sigma _2^3 \sigma _1+16 \sigma _3^3 \sigma _1\\
&&+20 \sigma _2 \sigma _3^2 \sigma _1+16 \sigma _2^2 \sigma _3 \sigma _1+18 \sigma _2^4+19 \sigma _3^4+3 \sigma _2 \sigma _3^3+11 \sigma _2^2 \sigma _3^2+9 \sigma _2^3 \sigma _3,\\
&&\sigma _1^4+19 \sigma _2 \sigma _1^3+20 \sigma _3 \sigma _1^3+12 \sigma _2^2 \sigma _1^2+19 \sigma _3^2 \sigma _1^2+22 \sigma _2 \sigma_3 \sigma_1^2+12 \sigma_2^3 \sigma _1+20 \sigma _3^3 \sigma _1\\
&&+10 \sigma _2 \sigma _3^2 \sigma _1+17 \sigma _2^2 \sigma _3 \sigma _1+5 \sigma _2^4+3 \sigma _3^4+11 \sigma _2 \sigma _3^3+17 \sigma _2^2 \sigma _3^2+22 \sigma _2^3 \sigma _3\rangle,
\end{eqnarray*}
\begin{eqnarray*}
\mathcal{I}_3^5&=&\langle6 \sigma _1^5+2 \sigma _2 \sigma _1^4+3 \sigma _3 \sigma _1^4+9 \sigma _2^2 \sigma _1^3+6 \sigma _3^2 \sigma _1^3+4 \sigma _2 \sigma _3 \sigma _1^3+3 \sigma _2^3 \sigma _1^2~~~~~~~~~~~~~~~~~~~~~~~~~~~~~~~~~~~~~~~~~~~~~~~\\
&&+12 \sigma _3^3 \sigma _1^2+12 \sigma _2 \sigma _3^2 \sigma _1^2+2 \sigma _2^2 \sigma _3 \sigma _1^2
+13 \sigma _2^4 \sigma _1+8 \sigma _3^4 \sigma _1+11 \sigma _2 \sigma _3^3 \sigma _1\\
&&+9 \sigma _2^2 \sigma _3^2 \sigma _1+4 \sigma _2^3 \sigma _3 \sigma _1+8 \sigma _2^5+7 \sigma _3^5+5 \sigma _2 \sigma _3^4+6 \sigma _2^2 \sigma _3^3+5 \sigma _2^3 \sigma _3^2,\\
&&7 \sigma _1^5+10 \sigma _2 \sigma _1^4+3 \sigma _3 \sigma _1^4+\sigma _2^2 \sigma _1^3+10 \sigma _3^2 \sigma _1^3+6 \sigma _2 \sigma _3 \sigma _1^3+13 \sigma _2^3 \sigma _1^2\\
&&+13 \sigma _3^3 \sigma _1^2+9 \sigma _2 \sigma _3^2 \sigma _1^2+4 \sigma _2^2 \sigma _3 \sigma _1^2+5 \sigma _2^5+\sigma _2^4 \sigma _1+13 \sigma _3^4 \sigma _1+3 \sigma _2 \sigma _3^3 \sigma _1\\
&&+2 \sigma _2^2 \sigma _3^2 \sigma _1+9 \sigma _2^3 \sigma _3 \sigma _1+10 \sigma _3^5+10 \sigma _2 \sigma _3^4+13 \sigma _2^2 \sigma _3^3+11 \sigma _2^3 \sigma _3^2+8 \sigma _2^4 \sigma _3,\\
&&2 \sigma _1^5+10 \sigma _2 \sigma _1^4+4 \sigma _3 \sigma _1^4+\sigma _2^2 \sigma _1^3+2 \sigma _3^2 \sigma _1^3+4 \sigma _2 \sigma _3 \sigma _1^3+8 \sigma _2^3 \sigma _1^2\\
&&+8 \sigma _3^3 \sigma _1^2+3 \sigma _2 \sigma _3^2 \sigma _1^2+5 \sigma _2^4 \sigma _1+8 \sigma _3^4 \sigma _1+7 \sigma _2 \sigma _3^3 \sigma _1+9 \sigma _2^2 \sigma _3^2 \sigma _1\\
&&+13 \sigma _2^3 \sigma _3 \sigma _1+8 \sigma _2^5+12 \sigma _3^5+7 \sigma _2 \sigma _3^4+2 \sigma _2^3 \sigma _3^2+7 \sigma _2^4 \sigma _3\rangle,
\end{eqnarray*}
\begin{eqnarray*}
\mathcal{I}_3^6&=&\langle3 \sigma _1^6+4 \sigma _2 \sigma _1^5+2 \sigma _3 \sigma _1^5+5 \sigma _2^2 \sigma _1^4+2 \sigma _3^2 \sigma _1^4+4 \sigma _2 \sigma _3 \sigma _1^4+\sigma _2^3 \sigma _1^3+\sigma _3^3 \sigma _1^3+2 \sigma _2 \sigma _3^2 \sigma _1^3~~~~~~~~~~\\
&&+\sigma _2^2 \sigma _3 \sigma _1^3+2 \sigma _2^4 \sigma _1^2+3 \sigma _3^4 \sigma _1^2+5 \sigma _2 \sigma _3^3 \sigma _1^2+\sigma _2^2 \sigma _3^2 \sigma _1^2+\sigma _2^3 \sigma _3 \sigma _1^2+3 \sigma _2^5 \sigma _1+4 \sigma _2 \sigma _3^4 \sigma _1\\
&&+3 \sigma _2^2 \sigma _3^3 \sigma _1+\sigma _2^3 \sigma _3^2 \sigma _1+4 \sigma _2^4 \sigma _3 \sigma _1+3 \sigma _2^6+2 \sigma _3^6+3 \sigma _2 \sigma _3^5+4 \sigma _2^2 \sigma _3^4+3 \sigma _2^3 \sigma _3^3+3 \sigma _2^4 \sigma _3^2,\\
&&4 \sigma _1^6+5 \sigma _2^6+3 \sigma _3^6+2 \sigma _2^2 \sigma _1^4+4 \sigma _3^2 \sigma _1^4+\sigma _2 \sigma _3 \sigma _1^4+3 \sigma _2^3 \sigma _1^3+3 \sigma _2 \sigma _3^2 \sigma _1^3+4 \sigma _2^4 \sigma _1^2+2 \sigma _2 \sigma _3^3 \sigma _1^2\\
&&+2 \sigma _3^4 \sigma _1^2+2 \sigma _2^2 \sigma _3^2 \sigma _1^2+3 \sigma _2^3 \sigma _3 \sigma _1^2+3 \sigma _3^5 \sigma _1+\sigma _2 \sigma _3^4 \sigma _1+2 \sigma _2^4 \sigma _3 \sigma _1+2 \sigma _2^2 \sigma _3^4+4 \sigma _2^3 \sigma _3^3+3 \sigma _2^5 \sigma _3,\\
&&5 \sigma _1^6+2 \sigma _2 \sigma _1^5+5 \sigma _3 \sigma _1^5+\sigma _2^2 \sigma _1^4+5 \sigma _3^2 \sigma _1^4+4 \sigma _2^3 \sigma _1^3+\sigma _3^3 \sigma _1^3+4 \sigma _2 \sigma _3^2 \sigma _1^3+5 \sigma _2^2 \sigma _3 \sigma _1^3+3 \sigma _2 \sigma _3^3 \sigma _1^2\\
&&+4 \sigma _3^4 \sigma _1^2+2 \sigma _2^2 \sigma _3^2 \sigma _1^2+3 \sigma _2^3 \sigma _3 \sigma _1^2+2 \sigma _2^5 \sigma _1+5 \sigma _3^5 \sigma _1+4 \sigma _2^2 \sigma _3^3 \sigma _1+2 \sigma _2^3 \sigma _3^2 \sigma _1+5 \sigma _2^4 \sigma _3 \sigma _1+4 \sigma _2^6\\
&&+2 \sigma _3^6+2 \sigma _2 \sigma _3^5+\sigma _2^3 \sigma _3^3+5 \sigma _2^4 \sigma _3^2+4 \sigma _2^5 \sigma _3\rangle.
\end{eqnarray*}
\begin{table}
\caption{The residues evaluated numerically at the intersection points are shown in the second column while the ones calculated by the conjecture are shown in the last column. Since the conjecture bypasses solving the equations numerically, the results in the last column are semi-analytical. Although not obvious at all, the values in the second and the last columns match up to computational precision.}
\begin{center}
\begin{tabular}{|c|c|c|}\hline
Ideal&Numerical& Residue from the ansatz\\ \hline
$\mathcal{I}_3^4$&$-0.00196517$&$-\frac{338369835974858276439763339608916939488780049506461387859}{172234849446370037753709508680189823992399891128510899390423}$\\ \hline
$\mathcal{I}_3^5$&$1.33251$&$\frac{185305421348253037675212877549168515918001249681372449355363515406576810500662817}{139071748905554975365797175643454675160373469042145930039888932779199639222995099}$\\ \hline
$\mathcal{I}_3^6$&$593.819 $&$\frac{219551617079855111701999235055855939627650975665549069197122883934238537987536968925296142}{370553184196614950880443018799517897307360951546021926710875320859728913170901077938477}$\\ \hline
$\mathcal{I}_4^2$&$-0.0023375$&$-\frac{2153841252590208111}{921484430712569847077}$\\ \hline
$\mathcal{I}_5^2$&$-0.0112606$&~~~~~~~~~$-\frac{473520568018181185074562119595694532674773643}{42044737913776287413645978868169636186479087179}$~~~~~~~~~\\ \hline
$\mathcal{I}_6^2$&$-0.00895405$&~~~~~~~~~~~~~~~~~~~~~~$-{\frac{~~~~~~~~~~~~~~~~~~~~~~~~~~~~~~~~~\updown{519880741176657236666450793152822696600712134946666110128882345754}{659894526078333385784278736240681558611734904658085876275390766034}~~~~~~~~~~~~~~~~~~~~~~~~~~~~~~~~~}{\updown{5967246404862841666772950558667081567881828770924504194620658839691}{2630580897251182112405482125334207621804213585965755701639782720971}}}$~~~~~~~~~~~~~~~~~~~\\ \hline
\end{tabular}
\end{center}
\label{HomId}
\end{table}%
\begin{eqnarray*}
\mathcal{I}_4^2&=&
\langle4 \sigma _1^2+4 \sigma _2 \sigma _1+4 \sigma _3 \sigma _1+3 \sigma _4 \sigma _1+\sigma _2^2+5 \sigma _3^2+\sigma _4^2+2 \sigma _2 \sigma _3+2 \sigma _2 \sigma _4+\sigma _3 \sigma _4,~~~~~~~~~~~~~~~~~~~~~~~\\
&&\sigma _1^2+\sigma _2 \sigma _1+2 \sigma _3 \sigma _1+4 \sigma _4 \sigma _1+5 \sigma _2^2+\sigma _4^2+4 \sigma _2 \sigma _3+5 \sigma _2 \sigma _4+2 \sigma _3 \sigma _4,\\
&&4 \sigma _1^2+2 \sigma _2 \sigma _1+2 \sigma _3 \sigma _1+2 \sigma _4 \sigma _1+3 \sigma _2^2+4 \sigma _3^2+5 \sigma _4^2+4 \sigma _2 \sigma _3+2 \sigma _2 \sigma _4+3 \sigma _3 \sigma _4,\\
&&2 \sigma _1^2+\sigma _2 \sigma _1+3 \sigma _3 \sigma _1+5 \sigma _2^2+5 \sigma _3^2+5 \sigma _2 \sigma _3+2 \sigma _2 \sigma _4+5 \sigma _3 \sigma _4\rangle,
\end{eqnarray*}
\begin{eqnarray*}
\mathcal{I}_5^2&=&\langle5 \sigma _1^2+5 \sigma _2 \sigma _1+3 \sigma _3 \sigma _1+\sigma _4 \sigma _1+2 \sigma _5 \sigma _1+\sigma _2^2+5 \sigma _5^2+5 \sigma _2 \sigma _4+\sigma _3 \sigma _4+5 \sigma _4 \sigma _5,\\
&&2 \sigma _1^2+3 \sigma _2 \sigma _1+\sigma _3 \sigma _1+4 \sigma _4 \sigma _1+5 \sigma _5 \sigma _1+\sigma _2^2+3 \sigma _3^2+5 \sigma _4^2\\
&&+\sigma _5^2+3 \sigma _2 \sigma _3+5 \sigma _2 \sigma _4+5 \sigma _3 \sigma _4+2 \sigma _3 \sigma _5+3 \sigma _4 \sigma _5,\\
&&5 \sigma _1^2+\sigma _3 \sigma _1+3 \sigma _4 \sigma _1+4 \sigma _3^2+2 \sigma _4^2+2 \sigma _5^2+2 \sigma _2 \sigma _3+\sigma _3 \sigma _4+\sigma _2 \sigma _5+\sigma _3 \sigma _5+2 \sigma _4 \sigma _5,\\
&&\sigma _1^2+\sigma _2 \sigma _1+\sigma _3 \sigma _1+2 \sigma _4 \sigma _1+3 \sigma _5 \sigma _1+5 \sigma _3^2+5 \sigma _4^2+2 \sigma _2 \sigma _3\\
&&+3 \sigma _2 \sigma _4+4 \sigma _3 \sigma _4+4 \sigma _2 \sigma _5+5 \sigma _3 \sigma _5+4 \sigma _4 \sigma _5,\\
&&\sigma _2^2+5 \sigma _1 \sigma _2+4 \sigma _3 \sigma _2+3 \sigma _4 \sigma _2+2 \sigma _5 \sigma _2+3 \sigma _3^2+2 \sigma _4^2+4 \sigma _1 \sigma _3+2 \sigma _1 \sigma _4+3 \sigma _3 \sigma _4+\sigma _3 \sigma _5+5 \sigma _4 \sigma _5\rangle,
\end{eqnarray*}
\begin{eqnarray*}
\mathcal{I}_6^2&=&\langle\sigma _1^2+3 \sigma _2 \sigma _1+3 \sigma _3 \sigma _1+4 \sigma _5 \sigma _1+5 \sigma _6 \sigma _1+5 \sigma _2^2+\sigma _3^2+\sigma _5^2+4 \sigma _6^2+3 \sigma _2 \sigma _3~~~~~~~~~~~~~~~~~~~~~~~~~~~~~~~~~~~~~~~~~~\\
&&+3 \sigma _2 \sigma _4+5 \sigma _3 \sigma _4+4 \sigma _2 \sigma _5+5 \sigma _3 \sigma _5+\sigma _2 \sigma _6+\sigma _3 \sigma _6+2 \sigma _4 \sigma _6+3 \sigma _5 \sigma _6,\\
&&5 \sigma _1^2+3 \sigma _2 \sigma _1+3 \sigma _4 \sigma _1+3 \sigma _5 \sigma _1+3 \sigma _6 \sigma _1+4 \sigma _2^2+3 \sigma _3^2+3 \sigma _4^2+5 \sigma _5^2+\sigma _6^2+5 \sigma _2 \sigma _4\\
&&+\sigma _3 \sigma _4+3 \sigma _2 \sigma _5+4 \sigma _3 \sigma _5+5 \sigma _4 \sigma _5+3 \sigma _2 \sigma _6+3 \sigma _3 \sigma _6+2 \sigma _4 \sigma _6+2 \sigma _5 \sigma _6,\\
&&5 \sigma _1^2+\sigma _2 \sigma _1+5 \sigma _3 \sigma _1+\sigma _4 \sigma _1+2 \sigma _5 \sigma _1+\sigma _6 \sigma _1+4 \sigma _2^2+5 \sigma _5^2+3 \sigma _6^2+\sigma _2 \sigma _3\\
&&+4 \sigma _2 \sigma _4+2 \sigma _3 \sigma _4+3 \sigma _2 \sigma _5+5 \sigma _3 \sigma _5+5 \sigma _2 \sigma _6+3 \sigma _3 \sigma _6+4 \sigma _4 \sigma _6,\\
&&2 \sigma _1^2+3 \sigma _2 \sigma _1+\sigma _3 \sigma _1+3 \sigma _4 \sigma _1+\sigma _5 \sigma _1+3 \sigma _6 \sigma _1+4 \sigma _2^2+5 \sigma _3^2+\sigma _4^2+5 \sigma _5^2+5 \sigma _6^2\\
&&+3 \sigma _2 \sigma _3+4 \sigma _2 \sigma _4+2 \sigma _3 \sigma _4+5 \sigma _2 \sigma _5+3 \sigma _3 \sigma _5+5 \sigma _4 \sigma _5+2 \sigma _3 \sigma _6+4 \sigma _4 \sigma _6+3 \sigma _5 \sigma _6,\\
&&5 \sigma _1^2+5 \sigma _3 \sigma _1+4 \sigma _4 \sigma _1+3 \sigma _5 \sigma _1+3 \sigma _6 \sigma _1+5 \sigma _2^2+3 \sigma _3^2+2 \sigma _4^2+4 \sigma _5^2+\sigma _6^2+3 \sigma _2 \sigma _3\\
&&+5 \sigma _2 \sigma _4+\sigma _3 \sigma _4+3 \sigma _2 \sigma _5+4 \sigma _3 \sigma _5+4 \sigma _4 \sigma _5+4 \sigma _2 \sigma _6+5 \sigma _3 \sigma _6+3 \sigma _4 \sigma _6+5 \sigma _5 \sigma _6,\\
&&3 \sigma _2^2+3 \sigma _3 \sigma _2+2 \sigma _4 \sigma _2+5 \sigma _5 \sigma _2+5 \sigma _3^2+2 \sigma _4^2+5 \sigma _5^2+\sigma _6^2+5 \sigma _1 \sigma _3+5 \sigma _1 \sigma _4\\
&&+5 \sigma _3 \sigma _4+5 \sigma _1 \sigma _5+3 \sigma _3 \sigma _5+5 \sigma _4 \sigma _5+4 \sigma _1 \sigma _6+3 \sigma _3 \sigma _6+\sigma _4 \sigma _6+2 \sigma _5 \sigma _6\rangle.
\end{eqnarray*}

\subsection{Inhomogeneous ideals} 
We have also checked Conjecture \ref{conj:2} against non-homogeneous ideals. This case, however, is much more time-consuming to work out and we only present four simple examples below that can be easily processed in a short period of time. The function $g(\sigma_i)$ remains the same as in the homogeneous case and the residues are listed in Table \ref{NonHomId}.
\begin{eqnarray*}
\mathcal{I}_3^{2,3}&=&
\langle2 \sigma _1^3+\sigma _3 \sigma _1^2+\sigma _1^2+2 \sigma _2 \sigma _1+\sigma _3 \sigma _1+\sigma _2^2+2 \sigma _2 \sigma _3^2+2 \sigma _3^2+2 \sigma _2^2 \sigma _3+2 \sigma _2 \sigma _3,\\
&&2 \sigma _1^3+2 \sigma _2 \sigma _1^2+2 \sigma _3 \sigma _1^2+\sigma _2^2 \sigma _1+\sigma _3^2 \sigma _1+2 \sigma _2^2+\sigma _3^2+\sigma _2 \sigma _3,\\
&&2 \sigma _1^3+\sigma _2^2 \sigma _1+2 \sigma _3^2 \sigma _1+2 \sigma _2 \sigma _1+\sigma _2 \sigma _3 \sigma _1+\sigma _2^3+\sigma _3^3+\sigma _2^2+\sigma _2^2 \sigma _3+2 \sigma _2 \sigma _3\rangle, \\
\mathcal{I}_3^{2,4}&=&\langle2 \sigma _1^4+\sigma _1^2+\sigma _3^3 \sigma _1+2 \sigma _3^4+2 \sigma _3^3,\sigma _1^4+\sigma _1^3+2 \sigma _2 \sigma _1+2 \sigma _3^4+\sigma _2^2 \sigma _3^2+2 \sigma _3^2,\\
&&2 \sigma _1^4+\sigma _2 \sigma _1^2+2 \sigma _3 \sigma _1+\sigma _2 \sigma _3^3+2 \sigma _2^2\rangle,\\
\mathcal{I}_3^{3,4}&=&\langle3 \sigma _1^3+\sigma _2 \sigma _3 \sigma _1+3 \sigma _2^2 \sigma _3^2+3 \sigma _2 \sigma _3^2,\sigma _1^4+5 \sigma _2 \sigma _1^2+5 \sigma _2^3+\sigma _2 \sigma _3^3,3 \sigma _2 \sigma _3^3+5 \sigma _3^3\rangle,\\
\mathcal{I}_4^{2,3}&=&\langle2 \sigma _1^2+5 \sigma _2 \sigma _1+2 \sigma _3 \sigma _4^2,2 \sigma _2^2+3 \sigma _4^2 \sigma _2+3 \sigma _1 \sigma _4,3 \sigma _3^2+\sigma _1 \sigma _3+2 \sigma _1 \sigma _4 \sigma _3,2 \sigma _1 \sigma _4+2 \sigma _2 \sigma _3 \sigma _4+5 \sigma _4 \sigma _4\rangle.
\end{eqnarray*}

\begin{table}
\caption{Non-homogeneous ideals}
\begin{center}
\begin{tabular}{|c|c|c|}\hline
Ideal&Numerical& Residue from the ansatz\\ \hline
$\mathcal{I}_3^{2,3}$&$-37.9119$&$-{859352384\over 22667121}$\\ \hline
$\mathcal{I}_3^{2,4}$&$-0.249991$&$-\frac{1}{4}$\\ \hline
$\mathcal{I}_3^{3,4}$&$0.837891$&$\frac{13121}{15625}$\\ \hline
$\mathcal{I}_4^{2,3}$&$0.102539$&${167007145709\over 1630641375000}$\\ \hline
\end{tabular}
\end{center}
\label{NonHomId}
\end{table}%

\section{Standard method for dealing with poles at infinity}
\label{STInfity}
Here we present a standard method of calculating the residue at infinity for the following differential form
\begin{align}\label{eq: form_at_infinity}
\Omega=\left( \frac{\sigma_2 \sigma_3^2}{h_1 h_2 h_3}-\frac{\sigma_1 \sigma_2 \sigma_3}{h_1 h_2 h_3} \right) d\sigma_1\wedge d\sigma_2 \wedge d\sigma_3
\end{align}
where $h_i$'s are the scattering equations in the gauge 
$\sigma_4=1$
\begin{align}
&h_1=l_4 + l_1 \sigma_1+l_2\sigma_2 + l_3 \sigma_3,\\
&h_2=-l_{14} \sigma_1-l_{24}\sigma_2 - l_{34} \sigma_3-l_{12}\sigma_1 \sigma_2-l_{23}\sigma_2 \sigma_3-l_{13} \sigma_1\sigma_3,\\
&h_3=l_{124}\sigma_1 \sigma_2+l_{234}\sigma_2 \sigma_3+l_{134} \sigma_1\sigma_3+l_{123}\sigma_1 \sigma_2 \sigma_3.
\end{align}

Let  us first recall  how to calculate the residue at infinity 
in the single variable case. In that case we have, after changing variable $z\rightarrow 1/\xi$,
\begin{align}
\oint\limits_{z=\infty} f(z)dz=\oint\limits_{\xi=0} f\left(\frac{1}{\xi}\right)d\left(\frac{1}{\xi}\right).
\end{align}
In the language of differential geometry we are actually considering the complex plane as one of two standard patches $U_0$ and $U_1$ covering $\mathbb{CP}^1$, i.e $U_0=\{[z_0,z_1]|z_0\neq 0\}$ and  $U_1=\{[z_0,z_1]|z_1\neq 0\}$, where $z_{0,1}$ are homogenous coordinates for $\mathbb{CP}^1$. Then the point of infinity is just the single point of $\mathbb{CP}^1$ that is missing in $U_0$, i.e $\infty=[0,1]\in \mathbb{CP}^1$. It is not on $U_0$ but on $U_1$, and the change of variable $z\rightarrow 1/\xi$ is just the coordinate transformation when we go from patch $U_0$ to $U_1$ where we can calculate the residue.

Similarly we can define the residue at infinity for the multivariable case. But in this case there is actually a hypersurface, instead of a single point, located at infinity. 
This can be seen as follows. 
To be specific  we discuss the calculation in the
form of~(\ref{eq: form_at_infinity}). 
Thus we are considering the form $\Omega$ on $\mathbb{C}^3$. Firstly we need to embed $\mathbb{C}^3$ into a compact 
manifold to be able to use the global residue theorem. 
 $\mathbb{CP}^3$ is a natural choice. The original 
 $\mathbb{C}^3$ can be identified with one of the standard 
 patches covering $\mathbb{CP}^3$, 
 say $U_0=\{[z_0,z_1,z_2,z_3]|z_0\neq 0\}$. 
 In that sense, what is now at infinity is the hypersurface 
 $\{[z_0,z_1,z_2,z_3]|z_0=0\}$. 
 Eq.~(\ref{eq: form_at_infinity}) is now interpreted 
  as the local expression on $U_0$ of a form on 
  $\mathbb{CP}^3$, i.e in terms of the homogenous coordinates,
\begin{align}
\Omega=\frac{(z_3/z_0-z_1/z_0)(z_2/z_0)(z_3/z_0)}{h_1h_2h_3}d\left(\frac{z_1}{z_0}\right)\wedge d\left(\frac{z_2}{z_0}\right)\wedge d\left(\frac{z_3}{z_0}\right)~.
\end{align}
And   $h_i$'s are expressed in terms of homogenous 
coordinates as well
\begin{align}
&h_1=(l_4 z_0+ l_1 z_1+l_2z_2 + l_3 z_3)z_0^{-1}=:\tilde{h}_1z_0^{-1},\\
&h_2=(-l_{14} z_0z_1-l_{24}z_0z_2 - l_{34} z_0z_3-l_{12}z_1 z_2-l_{23}z_2 z_3-l_{13} z_1z_3)z_0^{-2}=:\tilde{h}_2 z_0^{-2},\\
&h_3=(l_{124}z_0z_1 z_2+l_{234}z_0z_2 z_3+l_{134} z_0z_1z_3+l_{123}z_1 z_2 z_3)z_0^{-3}=:\tilde{h}_3 z_0^{-3}.
\end{align}
Furthermore the $\tilde{h}_i,~i=1,2,3$ as defined above  are 
homogenous functions of degrees $1,2,3$ respectively.
It follows that 
\begin{align}
\Omega=&\frac{(z_3-z_1)z_2 z_3}{\tilde{h}_1\tilde{h}_2\tilde{h}_3 z_0}\\\nonumber
&\times(z_0 dz_1\wedge dz_2\wedge dz_3-z_1dz_0\wedge dz_2\wedge dz_3-z_2dz_1\wedge dz_0\wedge dz_3-z_3 dz_1\wedge dz_2\wedge dz_0)~,
\end{align}
from which  $\Omega$ is singular on $\tilde{h}_1=0$, $\tilde{h}_2=0$, $\tilde{h}_3=0$ and $z_0=0$. To define residues on $\mathbb{CP}^3$ we need to regard the denominator $\tilde{h}_1\tilde{h}_2\tilde{h}_3 z_0$ as the product of three divisors. There are multiple ways to do so, each of which leads to an equation of global residue theorem. One choice, however, is particularly simple, i.e
\begin{align}
&D_1=\tilde{h}_1 z_0,\nonumber\\\label{eq: method1_divisors}
&D_2=\tilde{h}_2,\\\nonumber
&D_3=\tilde{h}_3.
\end{align}
With this choice the global residue theorem reads, with $S$ denoting the set of common zeros of these divisors,
\begin{align}\label{eq: GRT_method1}
0=\sum_{p\in S}\text{Res}_{\{D_1,D_2,D_3\},p}~\Omega=\sum_{p\in S\cap U_0}\text{Res}_{\{D_1,D_2,D_3\},p}~\Omega+\sum_{p\in S_\infty}\text{Res}_{\{D_1,D_2,D_3\},p}~\Omega
\end{align}
The first term is what we originally want to calculate, i.e the sum of all the residues \emph{not} at infinity. The second term contains contributions from points at infinity. $S_\infty$ denotes the set of poles at infinity, and is easily seen to be
\begin{align}
S_\infty=\{p_1:=[0,1,0,0],~p_2:=[0,0,1,0],~p_3:=[0,0,0,1]\}
\end{align}
whose elements are on $U_1$, $U_2$ and $U_3$ respectively. We now go on to each of these three patches to compute residues there. 

On $U_1$ we set $z_1=1$ and the form $\Omega$ is
\begin{align}
U_1:~~\left.\Omega\right|_{U_1}=-\frac{(z_3-1)z_2 z_3}{z_0\cdot \left.\left(\tilde{h}_1\tilde{h}_2\tilde{h}_3\right)\right|_{z_1=1}}dz_0\wedge dz_2\wedge dz_3.
\end{align}
The divisor choice (\ref{eq: method1_divisors}) becomes\footnote{Since $\tilde{h}_1(z_1=1,z_1=z_2=z_3=0)\neq 0$ the divisors are actually $\left.\{z_0,\tilde{h}_2,\tilde{h}_2,\tilde{h}_3\}\right|_{z_0=1}$.}
\begin{align}
U_1:~~D_1=z_0 \tilde{h}_1(z_1=1),~D_2=\tilde{h}_2(z_1=1),~D_3=\tilde{h}_3(z_1=1).
\end{align}
Thus
\begin{align}
\oint\limits_{p_1} \left.\Omega\right|_{U_1}&=\oint\limits_{z_0=z_2=z_3=0}\left[-\frac{(z_3-1)z_2 z_3}{z_0\cdot \left.\left(\tilde{h}_1\tilde{h}_2\tilde{h}_3\right)\right|_{z_1=1}}dz_0\wedge dz_2\wedge dz_3\right]=0\label{eq: res_U1}
\end{align}
and the residue at $p_1$ vanishes. 

On $U_2$ we set $z_2=1$ in $\Omega$ and choose the divisors in the same fashion, and the residue at $p_2$ also vanishes
\begin{align}
\oint\limits_{p_2}\left.\Omega\right|_{U_2}
&=\oint\limits_{z_0=z_1=z_3=0}\left[-\frac{(z_3-z_1)z_3}{z_0\cdot \left.\left(\tilde{h}_1\tilde{h}_2\tilde{h}_3\right)\right|_{z_2=1}}dz_0\wedge dz_2\wedge dz_3\right]=0.
\label{eq: res_U2}
\end{align}

The residue at $p_3$ is nontrivial. On $U_3$ we set $z_3=1$ and have
\begin{align}\nonumber
\oint\limits_{p_3}\left.\Omega\right|_{U_3}
&=\oint\limits_{z_0=z_1=z_3=0}\left[-\frac{(1-z_1)z_2}{z_0\cdot \left.\left(\tilde{h}_1\tilde{h}_2\tilde{h}_3\right)\right|_{z_2=1}}dz_1\wedge dz_2\wedge dz_0\right]\\\label{eq: res_U3}
&=-(l_{123}\cdot l_{23}\cdot l_{3})^{-1}.
\end{align}
Combining (\ref{eq: res_U1}), (\ref{eq: res_U2}) and (\ref{eq: res_U3}) we get 
\begin{align}
\sum_{p\in S_\infty}\text{Res}_{\{D_1,D_2,D_3\},p}~\Omega=-(l_{123}\cdot l_{23}\cdot l_{3})^{-1}.
\end{align}
Using momentum conservation, this is equal to
\begin{align}
\frac{1}{l\cdot k_3\left( -k_{23}+l\cdot k_2+l\cdot k_3 \right)\left(l\cdot k_4 \right)}.
\end{align}
By (\ref{eq: GRT_method1}) we then immediately get the value of 
$\sum_{p\in S\cap U_0}\text{Res}_{\{D_1,D_2,D_3\},p}~\Omega$, 
which agrees with (\ref{eq: res_result_method2}) found using our ansatz.


\section{Obtaining the residue by inspection}
As shown in Section \ref{sec:fourpoint}, we can evaluate the CHY-form conveniently using our ansatz. This method in fact applies to all kinds of meromorphic differential forms, and the intrinsic structure of the scattering equations has not been fully explored in calculations. From the discussion on residues at infinity in section \ref{STInfity}, it seems the particular form of scattering equations in fact greatly simplifies the evaluation process of the standard method. A natural question is then whether that could also help simplify the calculation using the ansatz method. As an example, in this section we discuss the following term from the 4-point one-loop super Yang-Mills amplitude
\begin{eqnarray}\label{eqn:I4part1re}
\oint\limits_{h_1=h_2=h_3=0}  \frac{d\sigma_1 \wedge d\sigma_2 \wedge d\sigma_3}{h_1 h_2 h_3}\, \mathcal R_4^{(1)} &=& - \oint\limits_{\tilde h_0=\cdots=\tilde h_3=0}  \, { d\sigma_1 \wedge d\sigma_2 \wedge d\sigma_3 \wedge d\sigma_0\over \tilde h_0 \tilde h_1\tilde h_2 \tilde h_3 }{\sigma_3 (\sigma_2-1)\over (\sigma_0-1)} \,.~~~~~~~
\end{eqnarray} 
Many equations from the local duality theorem contain only one of the $a_{ijkl}$'s. Such equations simply lead to the vanishing of those coefficients. The surviving equations from the local duality theorem are 
\begin{equation}
\left(
\begin{array}{ccccccccc}
 0 &  -l_{34} & 0 & -l_{23} & 0 & 0 & 0 & 0 & 0 \\
 0 & 0 & l_4 & 0 & l_2 & 0 & 0 & 0 & 0 \\
 l_3 & 0 &  l_2 & 0 & 0 & 0 & 0 & 0 & 3 l_4 \\
  l_4 &  l_3 & 0 & 0 & 0 & 0 & 0 & 0 & 0 \\
 0 &  l_4& 0 &  l_2 & 0 & 0 & 0 & 3 l_3 & 0 \\
 0 & 0 & 0 &  l_3 & 0 &  l_2& 0 & 0 & 0 \\
 0 & 0 & 0 & 0 &  l_4 &  l_3 & 3 l_2 & 0 & 0 \\
- l_{34}& 0 & -l_{24} & 0 & 0 & 0 & 0 & 0 & 0 \\
 0 & 0 & 0 & 0 &- l_{24} & -l_{23} & 0 & 0 & 0 \\
\end{array}
\right)
\left(\begin{array}{c}
a_{0012}\\a_{0021}\\a_{0102}\\a_{0120}\\a_{0201}\\a_{0210}\\a_{0300}\\a_{0030}\\a_{0003}
\end{array}
\right)=0.
\end{equation}
The equation for the intersection number is
\begin{eqnarray}
&&(-2 l_1 l_4 l_{12},2 l_1 l_3 l_{12},2 l_1 l_4 l_{13}, -2 l_1 l_3 l_{14},-2 l_1 l_2 l_{13},2 l_1 l_2 l_{14})\nonumber\\
&&\cdot (a_{0012},a_{0021},a_{0102},a_{0120},a_{0201},a_{0210})^T=6.
\label{Intsn}
\end{eqnarray}
From these equations it is possible to read off $a_{0012},a_{0021},a_{0102},a_{0120},a_{0201},a_{0210}$ by inspection
\begin{eqnarray}
&&(a_{0012},a_{0021},a_{0102},a_{0120},a_{0201},a_{0210})\nonumber\\
&=&({1\over 2 l_1 l_4 l_{34}}, {1\over -2 l_1 l_3 l_{34}}, {1\over -2 l_1 l_4 l_{24}}, {1\over 2 l_1 l_3 l_{23}}, {1\over 2 l_1 l_2 l_{24}}, {1\over-2 l_1 l_2 l_{23}}).
\end{eqnarray}
Namely the value of each $a$ is simply the reciprocal of the coefficient in front of it. We will see that this pattern also appears in later calculations for $\mathcal{R}_{4}^{(2)}$, $\mathcal{R}_{4}^{(3)}$ and $\mathcal{R}_{4}^{(4)}$. Now this knowledge is already enough for us to evaluate (\ref{eqn:I4part1re}) using our conjecture because
\begin{align}
\Res_{\{(\tilde h_1),(\tilde h_2),(\tilde h_3),(\tilde h_0)\}}\left({\sigma_3 (\sigma_2-1)\over (\sigma_0-1)}\right)=\mathbb{D} \left({ \sigma_3\sigma_2-\sigma_3\over \sigma_0-1}\right)=\mathbb{D} \left({-\sigma_3\over \sigma_0-1}\right)
\end{align} 
which involves only $a_{0012}$ and therefore
$$\Res_{\{(h_1),(h_2),(h_3)\}}(\mathcal R_4^{(1)})=2a_{0012}={1\over  l_1 l_4 l_{34}}.$$

The other three terms involving $\mathcal R_4^{(2)},\mathcal R_4^{(3)},\mathcal R_4^{(4)}$ can be calculated similarly. In evaluating the residue for $\mathcal R_4^{(2)}$, the intersection number equation is 
$$2 l_1 l_2  l_{13}a_{2010} -2 l_2 l_3 l_{13}a_{1020}  -2 l_2 l_4 l_{34}a_{0012} +2 l_2 l_3 l_{34}a_{0021} +2 l_2 l_4 l_{14}a_{1002} -2 l_1 l_2 l_{14} a_{2001} =6.$$
As mentioned above, the $a$'s are again reciprocals of their respective  coefficients
\begin{eqnarray}
&&(a_{2010},a_{1020},a_{0012},a_{0021},a_{1002},a_{2001})\nonumber\\
&=&({1\over2  l_1 l_2 l_{13}},{1\over-2  l_2 l_3 l_{13}}, {1\over-2 l_2 l_4 l_{34}},{1\over2  l_2 l_3 l_{34}},{1\over2  l_2 l_4 l_{14}},{1\over-2  l_1 l_2 l_{14}})
\end{eqnarray}
which at the same time solve other equations from local duality theorem. Thus 
$$\Res_{\{(h_1),(h_2),(h_3)\}}(\mathcal R_4^{(2)})=-2a_{2001}={1\over  l_1 l_2 l_{14}}.$$
For $\mathcal R_4^{(3)}$, The intersection equation is 
$$-2  l_2 l_3 l_{24}a_{0,2,0,1}+2  l_4 l_3 l_{24}a_{0102}-2 l_1 l_3 l_{12}a_{2100} +2  l_2 l_3 l_{12}a_{1200}+2  l_1 l_3 l_{14}a_{2001}-2  l_3 l_4 l_{14}a_{1002} =6,$$
from whose coefficients we find the solution to be
\begin{eqnarray}
&&(a_{0201},a_{0102},a_{2100},a_{1200},a_{2001},a_{1002})\nonumber\\
&=&({1\over -2  l_2 l_3 l_{24}},{1\over 2  l_4 l_3 l_{24}},{1\over -2 l_1 l_3 l_{12}}, {1\over 2  l_2 l_3 l_{12}},{1\over 2  l_1 l_3 l_{14}},{1\over -2  l_3 l_4 l_{14}}).
\end{eqnarray}
So we have
$$\Res_{\{(h_1),(h_2),(h_3)\}}(\mathcal R_4^{(3)})=2a_{1200}={1\over  -l_2 l_3 l_{12}}.$$
For $\mathcal R_4^{(4)}$, the intersection equation is 
$$2  l_4 l_3 l_{13}a_{1020}-2 l_1l_4 l_{13}a_{2010} -2   l_1 l_4 l_{12}a_{2100}-2  l_2 l_4 l_{12}a_{1200}+2  l_4 l_3 l_{23}a_{0120}+2  l_2l_4 l_{23}a_{0210}=6.$$
And the solution is 
\begin{eqnarray}
&&(a_{1020},a_{2010} ,a_{2100} ,a_{1200},a_{0120},a_{0210})\nonumber\\
&=&({1\over 2  l_4 l_3 l_{13}},{1\over-2 l_1l_4 l_{13}},{1\over-2  l_1 l_4 l_{12}},{1\over-2 l_2 l_4 l_{12}},{1\over-2  l_3 l_4 l_{23}},{1\over2  l_2l_4 l_{23}}).
\end{eqnarray}
Thus $$\Res_{\{(h_1),(h_2),(h_3)\}}(\mathcal R_4^{(4)})=-2a_{0120}={1\over   l_3 l_4 l_{23}}.$$

We expect such ease of finding the ansatz solution to appear also for higher point one-loop amplitudes, and will discuss it in future projects.
\acknowledgments
GC and TW thank Y. Zhang, E. Y. Yuan,  R.J. Huang, Y.H. Zhou,  K.J. Larsen, D. A. Kosower and H. Johansson for useful discussion and kind suggestions. GC, EC and TW  have been supported by  NSF of China Grant under Contract 11405084, the Open Project Program of State Key Laboratory of Theoretical Physics, Institute of Theoretical Physics, Chinese Academy of Sciences, China (No.Y5KF171CJ1). 

\bibliographystyle{JHEP}
\bibliography{ScatEq}

\providecommand{\href}[2]{#2}\begingroup\raggedright\begin{thebibliography}{10}

\bibitem{Cachazo:2013hca}
F.~Cachazo, S.~He, and E.~Y. Yuan, {\it {Scattering of Massless Particles in
  Arbitrary Dimensions}},  {\em Phys. Rev. Lett.} {\bf 113} (2014), no.~17
  171601, [\href{http://xxx.lanl.gov/abs/1307.2199}{{\tt arXiv:1307.2199}}].

\bibitem{Cachazo:2013iea}
F.~Cachazo, S.~He, and E.~Y. Yuan, {\it {Scattering of Massless Particles:
  Scalars, Gluons and Gravitons}},  {\em JHEP} {\bf 07} (2014) 033,
  [\href{http://xxx.lanl.gov/abs/1309.0885}{{\tt arXiv:1309.0885}}].

\bibitem{Cachazo:2014nsa}
F.~Cachazo, S.~He, and E.~Y. Yuan, {\it {Einstein-Yang-Mills Scattering
  Amplitudes From Scattering Equations}},  {\em JHEP} {\bf 01} (2015) 121,
  [\href{http://xxx.lanl.gov/abs/1409.8256}{{\tt arXiv:1409.8256}}].

\bibitem{Kawai:1985xq}
H.~Kawai, D.~C. Lewellen, and S.~H.~H. Tye, {\it {A Relation Between Tree
  Amplitudes of Closed and Open Strings}},  {\em Nucl. Phys.} {\bf B269} (1986)
  1--23.

\bibitem{Cachazo:2013gna}
F.~Cachazo, S.~He, and E.~Y. Yuan, {\it {Scattering equations and
  Kawai-Lewellen-Tye orthogonality}},  {\em Phys. Rev.} {\bf D90} (2014), no.~6
  065001, [\href{http://xxx.lanl.gov/abs/1306.6575}{{\tt arXiv:1306.6575}}].

\bibitem{Kleiss:1988ne}
R.~Kleiss and H.~Kuijf, {\it {Multi - Gluon Cross-sections and Five Jet
  Production at Hadron Colliders}},  {\em Nucl. Phys.} {\bf B312} (1989)
  616--644.

\bibitem{Bern:2008qj}
Z.~Bern, J.~J.~M. Carrasco, and H.~Johansson, {\it {New Relations for
  Gauge-Theory Amplitudes}},  {\em Phys. Rev.} {\bf D78} (2008) 085011,
  [\href{http://xxx.lanl.gov/abs/0805.3993}{{\tt arXiv:0805.3993}}].

\bibitem{Cachazo:2014xea}
F.~Cachazo, S.~He, and E.~Y. Yuan, {\it {Scattering Equations and Matrices:
  From Einstein To Yang-Mills, DBI and NLSM}},  {\em JHEP} {\bf 07} (2015) 149,
  [\href{http://xxx.lanl.gov/abs/1412.3479}{{\tt arXiv:1412.3479}}].

\bibitem{Roiban:2004yf}
R.~Roiban, M.~Spradlin, and A.~Volovich, {\it {On the tree level S matrix of
  Yang-Mills theory}},  {\em Phys. Rev.} {\bf D70} (2004) 026009,
  [\href{http://xxx.lanl.gov/abs/hep-th/0403190}{{\tt hep-th/0403190}}].

\bibitem{Witten:2003nn}
E.~Witten, {\it {Perturbative gauge theory as a string theory in twistor
  space}},  {\em Commun. Math. Phys.} {\bf 252} (2004) 189--258,
  [\href{http://xxx.lanl.gov/abs/hep-th/0312171}{{\tt hep-th/0312171}}].

\bibitem{Cachazo:2012pz}
F.~Cachazo, L.~Mason, and D.~Skinner, {\it {Gravity in Twistor Space and its
  Grassmannian Formulation}},  {\em SIGMA} {\bf 10} (2014) 051,
  [\href{http://xxx.lanl.gov/abs/1207.4712}{{\tt arXiv:1207.4712}}].

\bibitem{Skinner:2013xp}
D.~Skinner, {\it {Twistor Strings for N=8 Supergravity}},
  \href{http://xxx.lanl.gov/abs/1301.0868}{{\tt arXiv:1301.0868}}.

\bibitem{Cachazo:2013iaa}
F.~Cachazo, S.~He, and E.~Y. Yuan, {\it {Scattering in Three Dimensions from
  Rational Maps}},  {\em JHEP} {\bf 10} (2013) 141,
  [\href{http://xxx.lanl.gov/abs/1306.2962}{{\tt arXiv:1306.2962}}].

\bibitem{Cachazo:2012da}
F.~Cachazo and Y.~Geyer, {\it {A 'Twistor String' Inspired Formula For
  Tree-Level Scattering Amplitudes in N=8 SUGRA}},
  \href{http://xxx.lanl.gov/abs/1206.6511}{{\tt arXiv:1206.6511}}.

\bibitem{Cachazo:2012kg}
F.~Cachazo and D.~Skinner, {\it {Gravity from Rational Curves in Twistor
  Space}},  {\em Phys. Rev. Lett.} {\bf 110} (2013), no.~16 161301,
  [\href{http://xxx.lanl.gov/abs/1207.0741}{{\tt arXiv:1207.0741}}].

\bibitem{Dolan:2014ega}
L.~Dolan and P.~Goddard, {\it {The Polynomial Form of the Scattering
  Equations}},  {\em JHEP} {\bf 07} (2014) 029,
  [\href{http://xxx.lanl.gov/abs/1402.7374}{{\tt arXiv:1402.7374}}].

\bibitem{He:2014wua}
Y.-H. He, C.~Matti, and C.~Sun, {\it {The Scattering Variety}},  {\em JHEP}
  {\bf 10} (2014) 135, [\href{http://xxx.lanl.gov/abs/1403.6833}{{\tt
  arXiv:1403.6833}}].

\bibitem{Kalousios:2013eca}
C.~Kalousios, {\it {Massless scattering at special kinematics as Jacobi
  polynomials}},  {\em J. Phys.} {\bf A47} (2014) 215402,
  [\href{http://xxx.lanl.gov/abs/1312.7743}{{\tt arXiv:1312.7743}}].

\bibitem{Weinzierl:2014vwa}
S.~Weinzierl, {\it {On the solutions of the scattering equations}},  {\em JHEP}
  {\bf 04} (2014) 092, [\href{http://xxx.lanl.gov/abs/1402.2516}{{\tt
  arXiv:1402.2516}}].

\bibitem{Lam:2014tga}
C.~S. Lam, {\it {Permutation Symmetry of the Scattering Equations}},  {\em
  Phys. Rev.} {\bf D91} (2015), no.~4 045019,
  [\href{http://xxx.lanl.gov/abs/1410.8184}{{\tt arXiv:1410.8184}}].

\bibitem{Du:2016blz}
Y.-j. Du, F.~Teng, and Y.-s. Wu, {\it {CHY formula and MHV amplitudes}},  {\em
  JHEP} {\bf 05} (2016) 086, [\href{http://xxx.lanl.gov/abs/1603.08158}{{\tt
  arXiv:1603.08158}}].

\bibitem{Kalousios:2015fya}
C.~Kalousios, {\it {Scattering equations, generating functions and all massless
  five point tree amplitudes}},  {\em JHEP} {\bf 05} (2015) 054,
  [\href{http://xxx.lanl.gov/abs/1502.07711}{{\tt arXiv:1502.07711}}].

\bibitem{Cardona:2015eba}
C.~Cardona and C.~Kalousios, {\it {Comments on the evaluation of massless
  scattering}},  {\em JHEP} {\bf 01} (2016) 178,
  [\href{http://xxx.lanl.gov/abs/1509.08908}{{\tt arXiv:1509.08908}}].

\bibitem{Cardona:2015ouc}
C.~Cardona and C.~Kalousios, {\it {Elimination and recursions in the scattering
  equations}},  {\em Phys. Lett.} {\bf B756} (2016) 180--187,
  [\href{http://xxx.lanl.gov/abs/1511.05915}{{\tt arXiv:1511.05915}}].

\bibitem{Dolan:2015iln}
L.~Dolan and P.~Goddard, {\it {General Solution of the Scattering Equations}},
  \href{http://xxx.lanl.gov/abs/1511.09441}{{\tt arXiv:1511.09441}}.

\bibitem{Huang:2015yka}
R.~Huang, J.~Rao, B.~Feng, and Y.-H. He, {\it {An Algebraic Approach to the
  Scattering Equations}},  {\em JHEP} {\bf 12} (2015) 056,
  [\href{http://xxx.lanl.gov/abs/1509.04483}{{\tt arXiv:1509.04483}}].

\bibitem{Sogaard:2015dba}
M.~Sogaard and Y.~Zhang, {\it {Scattering Equations and Global Duality of
  Residues}},  {\em Phys. Rev.} {\bf D93} (2016), no.~10 105009,
  [\href{http://xxx.lanl.gov/abs/1509.08897}{{\tt arXiv:1509.08897}}].

\bibitem{Bosma:2016ttj}
J.~Bosma, M.~S¿gaard, and Y.~Zhang, {\it {The Polynomial Form of the Scattering
  Equations is an H-Basis}},  {\em Phys. Rev.} {\bf D94} (2016), no.~4 041701,
  [\href{http://xxx.lanl.gov/abs/1605.08431}{{\tt arXiv:1605.08431}}].

\bibitem{Zlotnikov:2016wtk}
M.~Zlotnikov, {\it {Polynomial reduction and evaluation of tree- and loop-level
  CHY amplitudes}},  {\em JHEP} {\bf 08} (2016) 143,
  [\href{http://xxx.lanl.gov/abs/1605.08758}{{\tt arXiv:1605.08758}}].

\bibitem{Cachazo:2015nwa}
F.~Cachazo and H.~Gomez, {\it {Computation of Contour Integrals on ${\cal
  M}_{0,n}$}},  {\em JHEP} {\bf 04} (2016) 108,
  [\href{http://xxx.lanl.gov/abs/1505.03571}{{\tt arXiv:1505.03571}}].

\bibitem{Gomez:2016bmv}
H.~Gomez, {\it {$\Lambda$ scattering equations}},  {\em JHEP} {\bf 06} (2016)
  101, [\href{http://xxx.lanl.gov/abs/1604.05373}{{\tt arXiv:1604.05373}}].

\bibitem{Cardona:2016bpi}
C.~Cardona and H.~Gomez, {\it {Elliptic scattering equations}},  {\em JHEP}
  {\bf 06} (2016) 094, [\href{http://xxx.lanl.gov/abs/1605.01446}{{\tt
  arXiv:1605.01446}}].

\bibitem{Baadsgaard:2015voa}
C.~Baadsgaard, N.~E.~J. Bjerrum-Bohr, J.~L. Bourjaily, and P.~H. Damgaard, {\it
  {Integration Rules for Scattering Equations}},  {\em JHEP} {\bf 09} (2015)
  129, [\href{http://xxx.lanl.gov/abs/1506.06137}{{\tt arXiv:1506.06137}}].

\bibitem{Lam:2015sqb}
C.~S. Lam and Y.-P. Yao, {\it {Role of Mšbius constants and scattering
  functions in Cachazo-He-Yuan scalar amplitudes}},  {\em Phys. Rev.} {\bf D93}
  (2016), no.~10 105004, [\href{http://xxx.lanl.gov/abs/1512.05387}{{\tt
  arXiv:1512.05387}}].

\bibitem{Lam:2016tlk}
C.~S. Lam and Y.-P. Yao, {\it {Evaluation of the Cachazo-He-Yuan gauge
  amplitude}},  {\em Phys. Rev.} {\bf D93} (2016), no.~10 105008,
  [\href{http://xxx.lanl.gov/abs/1602.06419}{{\tt arXiv:1602.06419}}].

\bibitem{Baadsgaard:2015hia}
C.~Baadsgaard, N.~E.~J. Bjerrum-Bohr, J.~L. Bourjaily, P.~H. Damgaard, and
  B.~Feng, {\it {Integration Rules for Loop Scattering Equations}},  {\em JHEP}
  {\bf 11} (2015) 080, [\href{http://xxx.lanl.gov/abs/1508.03627}{{\tt
  arXiv:1508.03627}}].

\bibitem{Mafra:2016ltu}
C.~R. Mafra, {\it {Berends-Giele recursion for double-color-ordered
  amplitudes}},  {\em JHEP} {\bf 07} (2016) 080,
  [\href{http://xxx.lanl.gov/abs/1603.09731}{{\tt arXiv:1603.09731}}].

\bibitem{Huang:2016zzb}
R.~Huang, B.~Feng, M.-x. Luo, and C.-J. Zhu, {\it {Feynman Rules of
  Higher-order Poles in CHY Construction}},  {\em JHEP} {\bf 06} (2016) 013,
  [\href{http://xxx.lanl.gov/abs/1604.07314}{{\tt arXiv:1604.07314}}].

\bibitem{Bjerrum-Bohr:2016juj}
N.~E.~J. Bjerrum-Bohr, J.~L. Bourjaily, P.~H. Damgaard, and B.~Feng, {\it
  {Analytic Representations of Yang-Mills Amplitudes}},
  \href{http://xxx.lanl.gov/abs/1605.06501}{{\tt arXiv:1605.06501}}.

\bibitem{Cardona:2016gon}
C.~Cardona, B.~Feng, H.~Gomez, and R.~Huang, {\it {Cross-ratio Identities and
  Higher-order Poles of CHY-integrand}},
  \href{http://xxx.lanl.gov/abs/1606.00670}{{\tt arXiv:1606.00670}}.

\bibitem{Cachazo:2015aol}
F.~Cachazo, S.~He, and E.~Y. Yuan, {\it {One-Loop Corrections from Higher
  Dimensional Tree Amplitudes}},  {\em JHEP} {\bf 08} (2016) 008,
  [\href{http://xxx.lanl.gov/abs/1512.05001}{{\tt arXiv:1512.05001}}].

\bibitem{Feng:2016nrf}
B.~Feng, {\it {CHY-construction of Planar Loop Integrands of Cubic Scalar
  Theory}},  {\em JHEP} {\bf 05} (2016) 061,
  [\href{http://xxx.lanl.gov/abs/1601.05864}{{\tt arXiv:1601.05864}}].

\bibitem{Berkovits:2013xba}
N.~Berkovits, {\it {Infinite Tension Limit of the Pure Spinor Superstring}},
  {\em JHEP} {\bf 03} (2014) 017,
  [\href{http://xxx.lanl.gov/abs/1311.4156}{{\tt arXiv:1311.4156}}].

\bibitem{Mason:2013sva}
L.~Mason and D.~Skinner, {\it {Ambitwistor strings and the scattering
  equations}},  {\em JHEP} {\bf 07} (2014) 048,
  [\href{http://xxx.lanl.gov/abs/1311.2564}{{\tt arXiv:1311.2564}}].

\bibitem{Geyer:2014fka}
Y.~Geyer, A.~E. Lipstein, and L.~J. Mason, {\it {Ambitwistor Strings in Four
  Dimensions}},  {\em Phys. Rev. Lett.} {\bf 113} (2014), no.~8 081602,
  [\href{http://xxx.lanl.gov/abs/1404.6219}{{\tt arXiv:1404.6219}}].

\bibitem{Geyer:2015bja}
Y.~Geyer, L.~Mason, R.~Monteiro, and P.~Tourkine, {\it {Loop Integrands for
  Scattering Amplitudes from the Riemann Sphere}},  {\em Phys. Rev. Lett.} {\bf
  115} (2015), no.~12 121603, [\href{http://xxx.lanl.gov/abs/1507.00321}{{\tt
  arXiv:1507.00321}}].

\bibitem{Geyer:2015jch}
Y.~Geyer, L.~Mason, R.~Monteiro, and P.~Tourkine, {\it {One-loop amplitudes on
  the Riemann sphere}},  {\em JHEP} {\bf 03} (2016) 114,
  [\href{http://xxx.lanl.gov/abs/1511.06315}{{\tt arXiv:1511.06315}}].

\bibitem{Geyer:2016wjx}
Y.~Geyer, L.~Mason, R.~Monteiro, and P.~Tourkine, {\it {Two-Loop Scattering
  Amplitudes from the Riemann Sphere}},
  \href{http://xxx.lanl.gov/abs/1607.08887}{{\tt arXiv:1607.08887}}.

\bibitem{cox2006using}
D.~A. Cox, J.~Little, and D.~O'shea, {\em Using algebraic geometry}, vol.~185.
\newblock Springer Science \& Business Media, 2006.

\bibitem{hartshorne2013algebraic}
R.~Hartshorne, {\em Algebraic geometry}, vol.~52.
\newblock Springer Science \& Business Media, 2013.

\bibitem{griffiths2014principles}
P.~Griffiths and J.~Harris, {\em Principles of algebraic geometry}.
\newblock John Wiley \& Sons, 2014.

\bibitem{Gross:1989ge}
D.~J. Gross and J.~L. Manes, {\it {The High-energy Behavior of Open String
  Scattering}},  {\em Nucl. Phys.} {\bf B326} (1989) 73--107.

\bibitem{Bern:2005iz}
Z.~Bern, L.~J. Dixon, and V.~A. Smirnov, {\it {Iteration of planar amplitudes
  in maximally supersymmetric Yang-Mills theory at three loops and beyond}},
  {\em Phys. Rev.} {\bf D72} (2005) 085001,
  [\href{http://xxx.lanl.gov/abs/hep-th/0505205}{{\tt hep-th/0505205}}].

\bibitem{Arkani-Hamed:2014via}
N.~Arkani-Hamed, J.~L. Bourjaily, F.~Cachazo, and J.~Trnka, {\it {Singularity
  Structure of Maximally Supersymmetric Scattering Amplitudes}},  {\em Phys.
  Rev. Lett.} {\bf 113} (2014), no.~26 261603,
  [\href{http://xxx.lanl.gov/abs/1410.0354}{{\tt arXiv:1410.0354}}].

\bibitem{Franco:2015rma}
S.~Franco, D.~Galloni, B.~Penante, and C.~Wen, {\it {Non-Planar On-Shell
  Diagrams}},  {\em JHEP} {\bf 06} (2015) 199,
  [\href{http://xxx.lanl.gov/abs/1502.02034}{{\tt arXiv:1502.02034}}].

\bibitem{Chen:2015qna}
B.~Chen, G.~Chen, Y.-K.~E. Cheung, R.~Xie, and Y.~Xin, {\it {Top-forms of
  Leading Singularities in Nonplanar Multi-loop Amplitudes}},
  \href{http://xxx.lanl.gov/abs/1506.02880}{{\tt arXiv:1506.02880}}.

\bibitem{Chen:2015bnt}
B.~Chen, G.~Chen, Y.-K.~E. Cheung, R.~Xie, and Y.~Xin, {\it {Top-forms of
  Leading Singularities in Nonplanar Multi-loop Amplitudes}},
  \href{http://xxx.lanl.gov/abs/1507.03214}{{\tt arXiv:1507.03214}}.

\bibitem{Bourjaily:2016mnp}
J.~L. Bourjaily, S.~Franco, D.~Galloni, and C.~Wen, {\it {Stratifying On-Shell
  Cluster Varieties: the Geometry of Non-Planar On-Shell Diagrams}},
  \href{http://xxx.lanl.gov/abs/1607.01781}{{\tt arXiv:1607.01781}}.

\bibitem{Baadsgaard:2015twa}
C.~Baadsgaard, N.~E.~J. Bjerrum-Bohr, J.~L. Bourjaily, S.~Caron-Huot, P.~H.
  Damgaard, and B.~Feng, {\it {New Representations of the Perturbative
  S-Matrix}},  {\em Phys. Rev. Lett.} {\bf 116} (2016), no.~6 061601,
  [\href{http://xxx.lanl.gov/abs/1509.02169}{{\tt arXiv:1509.02169}}].

\bibitem{Huang:2015cwh}
R.~Huang, Q.~Jin, J.~Rao, K.~Zhou, and B.~Feng, {\it {The Q-cut Representation
  of One-loop Integrands and Unitarity Cut Method}},  {\em JHEP} {\bf 03}
  (2016) 057, [\href{http://xxx.lanl.gov/abs/1512.02860}{{\tt
  arXiv:1512.02860}}].

\bibitem{ebeling2007functions}
W.~Ebeling, {\em Functions of several complex variables and their
  singularities}, vol.~83.
\newblock American Mathematical Soc., 2007.

\bibitem{greuel2007introduction}
G.-M. Greuel, C.~Lossen, and E.~I. Shustin, {\em Introduction to singularities
  and deformations}.
\newblock Springer Science \& Business Media, 2007.

\end{thebibliography}\endgroup
\end{document}